\documentclass[journal]{IEEEtran}

\usepackage{booktabs}
\usepackage{threeparttable}
\newtheorem{mydef}{Definition}
\usepackage{algorithmic}
\usepackage{algorithm}
\usepackage{amsfonts,amssymb,amsmath}
\usepackage{graphicx}
\usepackage{amsmath}
\usepackage{subfigure}
\usepackage{multirow}
\usepackage{float}
\usepackage{bbm,bm}
\usepackage[colorlinks]{hyperref}
\usepackage{color}
\newcommand{\tabincell}[2]{\begin{tabular}{@{}#1@{}}#2\end{tabular}}

\begin{document}

\title{Time-Variant Graph Classification}

\author{Haishuai Wang, Jia Wu, Xingquan~Zhu, \IEEEmembership{Senior Member,~IEEE}, \\Yixin Chen, Chengqi~Zhang, \IEEEmembership{Senior Member,~IEEE}

\IEEEcompsocitemizethanks{
\IEEEcompsocthanksitem H. Wang is with Department of Computer Science and Engineering, Washington University in St. Louis, MO, USA. E-mail: haishuai.wang@student.uts.edu.au.
\IEEEcompsocthanksitem J. Wu is with Centre for Quantum Computation \& Intelligent Systems (QCIS), Faculty of Engineering \& Information Technology, University of Technology Sydney, Australia. E-mail: jia.wu@uts.edu.au.
\IEEEcompsocthanksitem X. Zhu is with Department of Computer \& Electrical Engineering and Computer Science,
Florida Atlantic University, USA. E-mail: xzhu3@fau.edu.
\IEEEcompsocthanksitem Y. Chen is with Department of Computer Science and Engineering, Washington University in St. Louis, MO, USA. E-mail:
\IEEEcompsocthanksitem C. Zhang is with Centre for Quantum Computation \& Intelligent Systems (QCIS), Faculty of Engineering \& Information Technology, University of Technology Sydney, Australia. E-mail: chengqi.zhang@uts.edu.au.
}}

\maketitle
\pagestyle{empty}  
\thispagestyle{empty} 

\begin{abstract}
Graphs are commonly used to represent objects, such as images and text, for pattern classification. In a dynamic world, an object may continuously evolve over time, and so does the graph extracted from the underlying object. These changes in graph structure with respect to the temporal order present a new  representation of the graph, in which an object corresponds to a set of time-variant graphs. In this paper, we formulate a novel time-variant graph classification task and propose a new graph feature, called a \emph{graph-shapelet pattern}, for learning and classifying time-variant graphs. Graph-shapelet patterns are compact and discriminative \emph{graph transformation subsequences}. A graph-shapelet pattern can be regarded as a graphical extension of a \emph{shapelet} -- a class of discriminative features designed for vector-based temporal data classification. To discover graph-shapelet patterns, we propose to convert a time-variant graph sequence into time-series data and use the discovered shapelets to find \emph{graph transformation subsequences} as graph-shapelet patterns. By converting each graph-shapelet pattern into a unique tokenized graph transformation sequence, we can measure the similarity between two graph-shapelet patterns and therefore classify time-variant graphs. Experiments on both synthetic and real-world data demonstrate the superior performance of the proposed algorithms.
\end{abstract}


\IEEEpeerreviewmaketitle

\section{Introduction}
Graph classification is an important branch of data mining research, given that much structured and semi-structured data can be represented as graphs. Images \cite{Yang:TSMCS09,deng2014weakly}, text \cite{Wu:TKDE14,Yuan:TSMCS15,Liu:TSMCS13}, fault diagnosis \cite{Sztyber:TSMCS15}, decision making \cite{Rego:TSMCS15} and biological data \cite{Deshpande:TKDE05,Shang:TSMCS15} present just a few examples. The main challenge in graph classification is that graphs do not have vectorial features directly available for classification, therefore traditional vector-based classifiers such as support vector machines (SVM) are not applicable. Research efforts in this area have focused on extracting discriminative substructures from graph data, so that graphs can be represented as vectorial features for learning and classification. One of the most widely used methods extracts frequent subgraph patterns as discriminative features, and many studies have shown that this approach achieves good performance \cite{Kong:KDD11,hongflickr,sabirin2012moving}.

\begin{figure}[!t]
  \centering
  \includegraphics[width=0.45\textwidth]{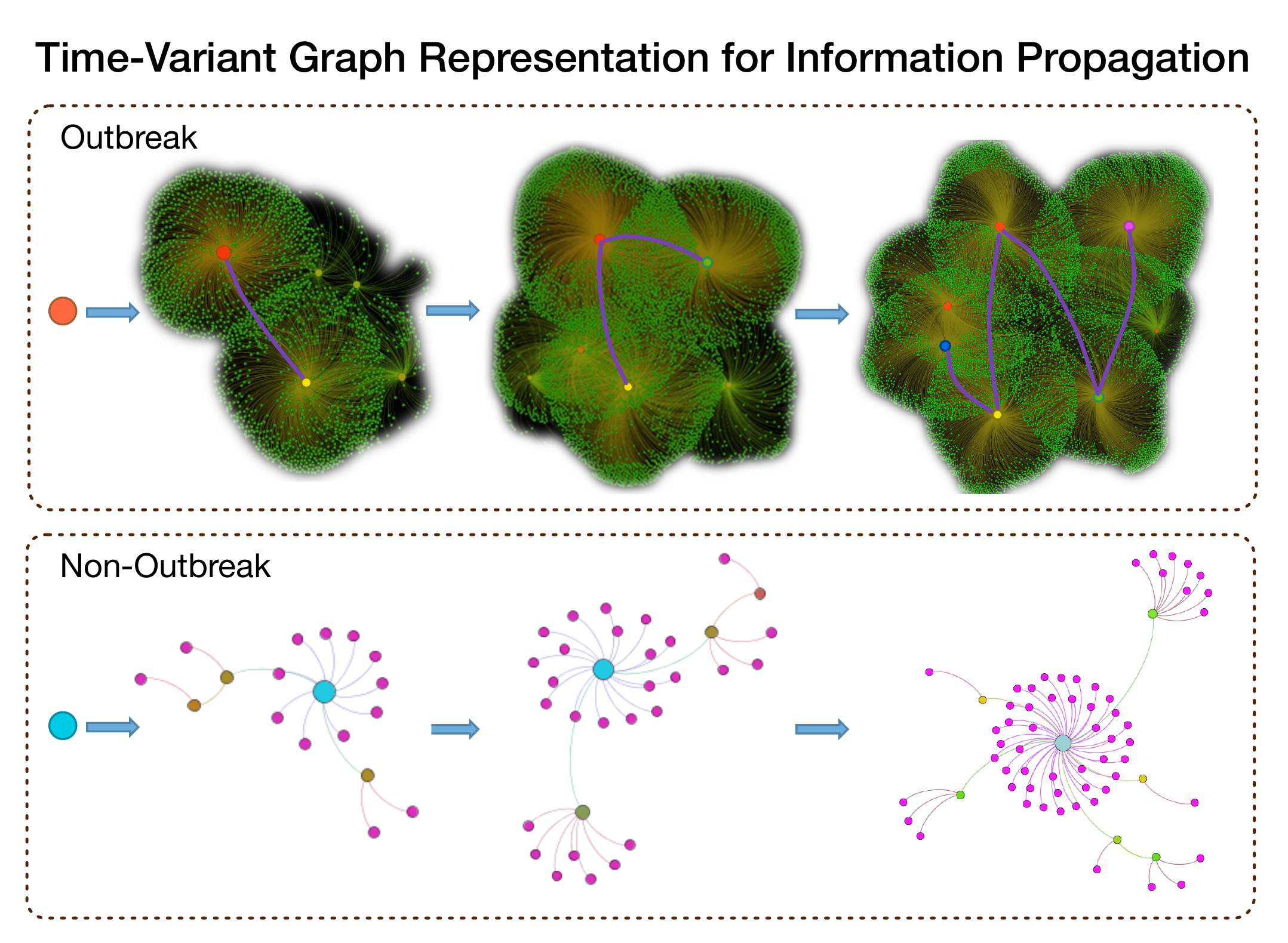}
  \caption{An example of a time-variant graph representation for large-scale information propagation. The information propagation can be regarded as a series of individual graphs. At different time periods, both the vertex volume and the graph structure are diverse, \textit{i.e.,} the information propagation is a process that takes place on the graph. The information propagation outbreak prediction aims to build a time-variant graph classification solution to accurately differentiate an outbreak time-variant graph (the entire row above) from a non-outbreak time-variant graph (the entire row below).}
  \label{fig:cascades}
\end{figure}

Despite positive results for many applications \cite{Kong:SDM13,wang2015defragging}, using discriminative subgraph pattern as features is based on a strict assumption that graph data are static and that frequent subgraph patterns are obtainable. However, time-variant graph data generated from dynamic graphs has recently entered the domain, for example, for the information propagation of graphs (\textit{i.e.} a series of graphs in time series). If the information propagation in time windows is analyzed, a sequence of graphs showing information propagation over time, where the target is to predict the class label that describes the outbreak or non-outbreak of an information propagation record, as shown in Fig. \ref{fig:cascades}. A sudden increase in occurrences of information propagated over social media in a particular time and place is known as outbreak information propagation  \cite{wang2017incremental}. In other words, outbreak occurs when information is propagated to a large audience in a very short time (\emph{e.g.}, a common case of outbreak is rumor or malicious information diffusion \cite{wang2017towards}). At a specific point in time, the status of the information diffusion is a graph, but over time the graph is diverse. Therefore, an information diffusion at different stages forms a time-variant graph. As both the vertex volumes and structure keep changing in each time-variant graph, outbreak information propagation is characterized by a number of special structures that can be modeled as graph-shapelet patterns.

In this paper, we refer to a sequence of graphs with a strict temporal order as a \textbf{time-variant graph}, \textit{e.g.,} an entire row in Fig. \ref{fig:cascades} or \ref{example_intro}. A time-variant graph can be used to characterize the changing nature of a structured object. For example, a chemical compound \cite{Deshpande:TKDE05} can be represented as a graph. By varying the temperature or other environmental conditions, the structure of the chemical compound with respect to each changed condition can be recorded, and a set of graphs (\textit{i.e.} a time-variant graph) will capture its evolution. This representation is much more comprehensive and more accurate than simply representing a chemical compound as a single static graph.

In time-variant graph classification, frequent subgraph patterns are very difficult to find because it is necessary to consider both the graph structure space and the temporal correlations to find subgraph candidates for validation. The search space becomes infinite and hence unlikely to yield stable structures. Secondly, the distance between subgraph patterns may change irregularly over time, therefore even though a relatively large threshold parameter could be set to mine frequent subgraph patterns, the resulting set may be empty.

Although there are many works on graph stream classification \cite{Aggarwal:SDM11,Pan:TCYB15}, these works typically assume that graphs are independent of one another and flow in a stream fashion. As a result, subgraph features are still applicable because the structure of each graph is static and the static graph distribution assumption and satisfactory empirical results have been reported~\cite{Seo:GIP2015}. However, in time-variant graph classification, the entire graph data changes over time and therefore obtaining stable subgraph features is unlikely. New types of graph patterns are needed as features for time-variant graph classification.

In this paper, we outline a new class of graph patterns, called \textbf{graph-shapelet patterns}, that can be used as features for time-variant graph classification.  Graph-shapelet patterns are inspired by the \emph{shapelets} in \cite{Ye:KDD09}, which were proposed as discriminative features for time-variant data stream classification. Shapelets are compact and discriminative subsequences that significantly reduce the time and space required for time-variant data classification \cite{Mueen:KDD11,Grabocka:KDD14}. However, all the existing time series issues fall into either univariate or multi-variate problems. They ignore the structural information behind time sequences that can be represented as graphs.

The design of graph-shapelet patterns requires a graph sequential pattern mining to be grafted onto time series shapelet pattern mining. The main technical challenges include:

\begin{itemize}
  \item \emph{Challenge \#1: } From the time series shapelet pattern mining viewpoint, existing shapelet pattern mining is only based on univariant/multi-variant time series data. Extending existing shapelet pattern mining algorithms to time-variant graphs is the first challenge, and requires downscaling a graph sequence into a univariant/multi-variant time series by designing a graph statistics, which can minimize information loss during the conversion of a graph into a univaraint/multivariant time series.

  \item \emph{Challenge \#2: } From the graph sequence mining viewpoint, existing graph sequence mining aims to find frequent transformation subsequences \emph{directly} from graph sequences by using \emph{graph edit similarity}, which measures the similarity between two neighboring graphs. However, such methods are unscalable to the large sets of graph sequences which commonly occur in dynamic graphs. Moreover, these graph sequence mining algorithms cannot always guarantee that the frequent graph transformation subsequences are \emph{discriminative transformation subsequences} for graph classification. Therefore, how to design a scalable algorithm that can extend existing graph sequence mining algorithms to discover \emph{discriminative transformation subsequences} is the second challenge.

  \item \emph{Challenge \#3: } From the application viewpoint, time-variant graphs have the potential to represent large and dynamic graph data. How to evaluate the performance of the proposed method by collecting persuasive data sets and designing proper benchmark methods is the third challenge.
\end{itemize}

In this paper, we propose a new class of graph patterns, \emph{graph-shapelet patterns}, as features for time-variant graph classification. Graph-shapelet patterns are compact and discriminative graph transformation subsequences that can be used to classify graph sequences. Technically, to solve these challenges, we first convert a graph sequence into a time series using fast graph statistics. Then, we extract shapelets from the converted time series by using a traditional shapelets mining algorithms. Next, we locate the graph subsequences that match the shapelets from the original graph sequences. Each \emph{graph subsequence} corresponds to a unique \emph{graph transformation subsequence} based on our new graph transformation rules. At the last step, we extract the most discriminative \emph{graph transformation subsequences} as graph-shapelets based on their graph edit similarity. Experiments on both synthetic and real-world data demonstrate the performance of the proposed algorithms.

The remainder of the paper is organized as follows. Related works are reviewed in Section \ref{relatedWork}. Section~\ref{pre} presents the preliminaries with key definitions. Section~\ref{tpm} introduces the proposed graph-shapelet pattern based classifier. Section~\ref{experiments} reports the experimental results, with discussion being conducted in Section \ref{sec:dis}. We conclude the paper in Section~\ref{con}.

\section{Related Work} \label{relatedWork}

The increasing availability of structured data holds great potential for knowledge discovery from graph data. Generally, real graphs tend to be time-variant graphs with time changes. However, one of the difficulties when classifying time-variant graphs is converting the graph data into a format that allows the extraction of effective features for graph classification.

Exploring new graph features is a persistent research focus in the graph data mining community, and many feature extraction methods have been proposed~\cite{Kong:SDM13,wang2015defragging}. Existing graph classification approaches can be roughly categorized into two groups: 1) distance based methods that include a pairwise similarity measure between two graphs, such as graph kernel~\cite{Bai:PR16}, graph embedding~\cite{Riesen:TSMCB09,Maronidis:PR15,Rossi:PR15,Ding:PR15}, graph matching \cite{Serratosa:PR15,sang2012robust}, and transformation~\cite{Izadi:TIP12,Li:TCADICS12,Wang:ICV15}; and 2) subgraph feature based methods that identify significant subgraphs as signatures for one particular class. For example, the work in~\cite{Fei:TKDD14} proposed to the extraction of subgraph structural information for classification. The work in~\cite{Thoma:SDM09} formulated the subgraph selection problem as a combinatorial optimization problem and used heuristic rules and a combination of frequent subgraph mining algorithms such as gSpan~\cite{Yan:ICDM02} to find the subgraph features. Some boosting methods, such as~\cite{Saigo:08ML,Pan:TKDE15_2} use individual subgraph features as weak classifiers to build an ensemble for graph classification, yet none of these methods consider the dynamic property of graphs.

Several graph stream mining methods have recently been proposed that address the streaming feature challenge. The work in \cite{Perkins:ICML03} proposed a grafting algorithm based on a stage-wise gradient descent approach for streaming feature selection. \cite{Zhou:JMLR06} studied streaming feature selection and proposed two novel algorithms based on streamwise regression, information investing and alpha investing. The work in \cite{Ungar:AISTATS05} developed a new, adaptive complexity penalty, the Information Investing Criterion (IIC), to the model to dynamically adjust the threshold of the entropy reduction required for adding a new feature. In \cite{Pan:TCYB15}, a graph ensemble boosting approach was proposed to handle the imbalance in noisy graph stream classification. Again, these graph streaming methods ignore the dynamic nature of graph sequences.

Discriminative features for temporal data classification have been studied extensively~\cite{Ding:PVLDB08, Keogh:DMKD03, Xi:ICML06}. Examples include bursts~\cite{Kleinberg:KDD02}, periods~\cite{Elfeky:TKDE05}, anomalies~\cite{Wei:SSDBM05}, motifs~\cite{Lin:KDD02}, shapelets~\cite{Ye:KDD09} and discords~\cite{Yankov:KDD07}. Time series shapelets have recently attracted increased interest in data mining ~\cite{Hartmann:SMC10, Mcgovern:DMKD11, Xing:SDM11}, because shapelets are usually much shorter than the original time series, which allows us to use a single shapelet for classification instead of the entire data set. Shapelets were first proposed by~\cite{Ye:KDD09} as time-series segments that maximally predict the target variable, however the runtime of brute-force shapelet discovery is not feasible due to the large number of candidates. A series of speed-boosting techniques, such as early abandonment of distance computations and entropy pruning of the information gain metric, have therefore been proposed~\cite{Ye:KDD09}. The work in \cite{Mueen:KDD11} relies on the reuse of computations and pruning of the search space to speed up the discovery of shapelets. \cite{Grabocka:KDD14} proposed a novel method which learns near-to-optimal shapelets directly, without the need to search exhaustively among a pool of candidates extracted from time-series segments. Nevertheless, all the time series issues addressed fall into either the univariate or multivariate problem category, and ignore the structural information behind a time series which could be represented by graphs.

\begin{figure*}[!t]
\centering
\includegraphics[width=0.8\textwidth]{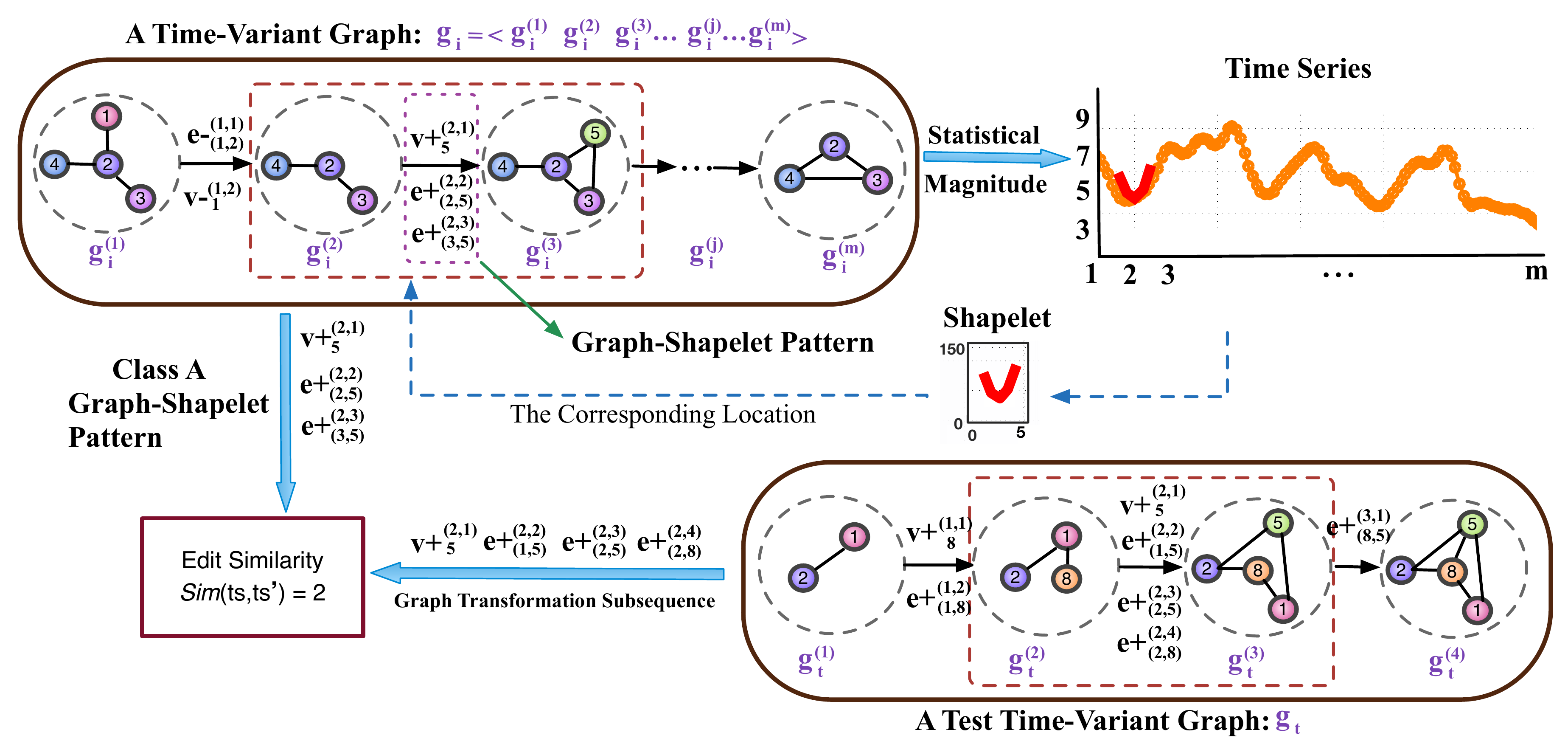}
\caption{A toy example of graph-shapelet pattern mining to explain the definitions of related operations. We first explore univariate time series from a set of time-variant graphs (detailed in Section~\ref{gs}), where shapelets are discovered using shapelet pattern mining algorithms. The graph-shapelet patterns (discriminative graph transformation sequences) are then relocated by calculating the graph edit similarity (as in Definition \ref{def:es}) between shapelet patterns which match a set of graph subsequences. To make the above illustration clear, we use this example through \emph{Examples} \protect\hyperlink{1}{1} $\sim$ \protect\hyperlink{4}{4}.}
\label{fig:example}
\end{figure*}

The work in \cite{Inokuchi:ICDM08} proposes a method to mine frequent subsequences from graph sequence data. However, our work is different. First, \cite{Inokuchi:ICDM08} focuses on frequent graph transformation sequence pattern mining rather than a classification problem. Therefore, it does not consider the label of each graph sequence. Second, it initially transfers the whole graph sequences to transformation sequences and then mines the frequent patterns. However, this method is only efficient for small time-variant graph data sets since the search space tends to be too large for large-scale data sets (such as outbreak information propagation). Instead of generating the transformation sequences from the whole data set directly, our method first finds discriminative graph subsequences (graph-shapelets) to reduce the transformation space. Finally, we use real-world information propagation to evaluate our proposed approach in the experimental section. We first model the information propagation as time-variant graph and then classify the outbreak diffusion from a non-outbreak diffusion. However, \cite{Inokuchi:ICDM08} is not applicable to this problem.

In summary, this work advances the state-of-the-art by combining existing research in graph sequence mining and time series shapelet mining to extract graph-shapelets for time-variant graph classification.

\section{Problem Formulation and Preliminaries}
\label{pre}

\subsection{Problem Definition}

\begin{mydef}{(\emph{Graph})}
For a time-variant graph $g_i$, each graph $g_i^{(j)}$ is represented as $g_i^{(j)} = (\mathcal{V}, E)$, where $\mathcal{V} = \{v_1, \cdots, v_z\}$ is a set of vertices, $E = \{(v,v^{\prime})|(v,v^{\prime}) \in \mathcal{V} \times \mathcal{V}\}$ is a set of edges. $\mathcal{V}(g_i^{(j)}), E(g_i^{(j)})$ are sets of vertices and edges of $g_i^{(j)}$.
\end{mydef}

\begin{mydef}{(\emph{Time-Variant Graph})}
A time-variant graph (\textit{i.e.,} graph sequence) is represented by $g_i = \langle g_i^{(1)} \cdots g_i^{(j)} \cdots g_i^{(m)}\rangle$, where the superscript represents the temporal order of the graph (\textit{e.g.,} $g_i^{(1)}$) in the sequence, and each time-variant $g_i$ is assigned a label $L_i$.
\end{mydef}

\begin{mydef}{(\emph{Sub-Time-Variant Graph})}
For a time-variant graph (\textit{i.e.,} graph sequence) $g_i = \langle g_i^{(1)} \cdots g_i^{(m)}\rangle$, a sub-time-variant graph $g_i^{\prime} = \langle g_i^{(k)} \cdots g_i^{(m')}\rangle$ is a fragment of a time-variant graph, where $1 \leq k \leq m' \leq m$.
\end{mydef}

Given a set of time-variant graphs $\mathcal{G}=\{g_1, \cdots, g_n\}$, time-variant graph classification \textbf{aims} to learn classification models from $\mathcal{G}$ to accurately predict previously unseen time-variant graphs with maximum accuracy. For simplicity, we only consider binary time-variant graph classification tasks. Since an information propagation graph sequence is a special case of a time-variant graph, we use information propagation as an example of a time-variant graph. Thus, the specific classification task is to predict an outbreak information propagation from a non-outbreak information propagation.

\subsection{Preliminaries}
In this subsection, we formally define notations and introduce time series data and graph transformation subsequences. We state the distance between the time-variant graphs and the two approaches to convert the time-variant graphs into graph transformation sequences. Fig.~\ref{fig:example} shows a toy example of the given definitions and the proposed approaches. The main operations used in the graph dynamic changes are defined in Table \ref{tab:admissibility}.

Since the proposed method is related to time series shapelets, and the shapelet is from time series segments, we first give some prior knowledge on time series segments and shapelets to better understand the proposed approach.

\emph{Time series segments.} Consider a sliding window of length $t$: a set of segments can be obtained when the window slides along a time series. For time series $\bm{x}_j \in \bm{X}$, we can generate a total of $q-t+1$ segments by sliding the window from $\bm{x}_j^{(1)}$ to $\bm{x}_j^{(q-t+1)}$. Thus, for the entire time series $\bm{X}$, there is a total of $(q-t+1) \times n$ segments, i.e., $\bm{\Omega} = [\varphi_1, \cdots, \varphi_{(q-t+1) \times n}]$ where each $\varphi_j \in \bm{\Omega}$ denotes a segment. Each element $\bm{s}_{(j, k)}$ is the distance between time series $\bm{x}_j$ and segment $\varphi_k$. It can be defined as the differential minimum distance that approximately denotes the minimum distance between the time series and the segment. Note that the segment length $t \ll q$, and the number of segments $(q-t+1) \times n$ is very large.

\emph{Shapelets.} Shapelets are defined as the most discriminative subsequences (segments) of time series that can best predict the target variable. Shapelets can capture inherent structures of time series, contributing to a high prediction of accuracy as explainable features \cite{Ye:KDD09}. Therefore, time series segments are shapelet candidates, and we can use $\bm{\Omega}$ as the feature space for shapelet selection. To represent each time series $\bm{x}_j \in \bm{X}$ in the space $\bm{\Omega}$, we use a column vector $\bm{s}_i = [s_{i,1}, \cdots, s_{(i, (q-t+1) \times n)}]$ to record $\bm{x}_j's$ feature values, where each element $s_{i,j}$ depends on a distance function between $\bm{x}_j$ and segment $\varphi_j \in \bm{\Omega}$, i.e., $s_{i,j} = d_{(\bm{x}_i, \varphi_j)}$. This way, the time series data set $\bm{X}$ can be represented by a data matrix $\bm{S} = [\bm{s}_1, \bm{s}_2, \cdots, \bm{s}_n] \in \mathbb{R}^{(q-t+1)\times n,n}$, where each column vector $\bm{s}_j$ represents a time series $\bm{x}_j$ in space $\bm{\Omega}$.

We propose the graph-shapelet by borrowing the idea of shapelets in the time series. The similarity between time series and time-variant graphs is that they are time-related sequential data, whereas the difference at each time point is numerical value in a time series and a graph in a time-variant graph. Therefore, we can calculate the Euclidian distance after learning about shapelets from the time series. In contrast, we can calculate the edit similarity after learning about graph-shapelets from time-variant graphs.

In a time-variant graph $g_i$, we assume that vertices $v$ and $v^{\prime}$ represent different objects (\emph{e.g.}, users) in $g_i^{(j)}$. We define a set of unique IDs $ID(\mathcal{V})$ and pairs of unique IDs $ID(E)$ as:
\begin{align*}
&ID(\mathcal{V}) = \{id(v)|v \in \cup_{g_i^{(j)}\in g_i}\mathcal{V}(g_i^{(j)})\},\\ &ID(E) = \{(id(v),id(v^{\prime}))|(v,v^{\prime}) \in \cup_{g_i^{(j)}\in g_i}E(g_i^{(j)})\}.
\end{align*}

\begin{table}[!t]
\addtolength{\tabcolsep}{-3.5pt}
\renewcommand{\arraystretch}{1.2}
\centering
\caption{Operation Definitions}
\label{tab:admissibility}
\begin{tabular}{|c|l|}
  \hline
  $v+^{(j,k)}_u$ & \tabincell{l}{$\neg \exists v \in \mathcal{V}(g_i^{(j)})~s.t.~id(v) = u$ and\\ $\exists v^{\prime} \in \mathcal{V}(g_i^{(j+1)})~s.t.~id(v^{\prime}) = u\}$}\\
  \hline
  $v-^{(j,k)}_u$ & \tabincell{l}{$\exists v \in \mathcal{V}(g_i^{k})~s.t.~id(v) = u$,\\$\neg\exists(v_1,v_2) \in E(g_i^{(k)})~s.t.~(id(v_1) = u) \vee (id(v_2) = u)$\\ and $\neg\exists v^{\prime} \in \mathcal{V}(g_i^{k+1})~s.t.~id(v^{\prime}) = u\}$} \\
  \hline
  $e+_{(u_1,u_2)}^{(j,k)}$ & \tabincell{l}{$\neg\exists(v_1,v_2) \in E(g_i^{(j)}), \exists v_1 \in \mathcal{V}(g_i^{(j)}),$ \\ $\exists v_2 \in \mathcal{V}(g_i^{(j)})~s.t. (id(v_1) = u_1) \wedge (id(v_2) = u_2)$ and \\ $\exists(v_1^{\prime}, v_2^{\prime}) \in E(g_i^{(j+1)}) s.t. (id(v_1^{\prime} = u_1) \wedge (id(v_2^{\prime} = u_2)$\\}  \\
  \hline
  $e-_{(u_1,u_2)}^{(j,k)}$ & \tabincell{l}{$\neg\exists(v_1,v_2) \in E(g_i^{(j)}), s.t. (id(v_1) = u_1) \wedge (id(v_2) = u_2)$ \\ and $\neg\exists(v_1^{\prime}, v_2^{\prime}) \in E(g_i^{(j+1)}), \exists v_1^{\prime} \in \mathcal{V}(g_i^{j+1}),$ \\ $\exists v_2^{\prime} \in \mathcal{V}(g_i^{(j+1)})~s.t. (id(v_1^{\prime} = u_1) \wedge (id(v_2^{\prime} = u_2)$}  \\
  \hline
\end{tabular}
\begin{tablenotes}
\item[] ~~~$*$:~$k$ is the index of operation in a transformation sequence.
\end{tablenotes}
\end{table}

\begin{figure*}[!t]
\centering
\includegraphics[width=0.95\textwidth]{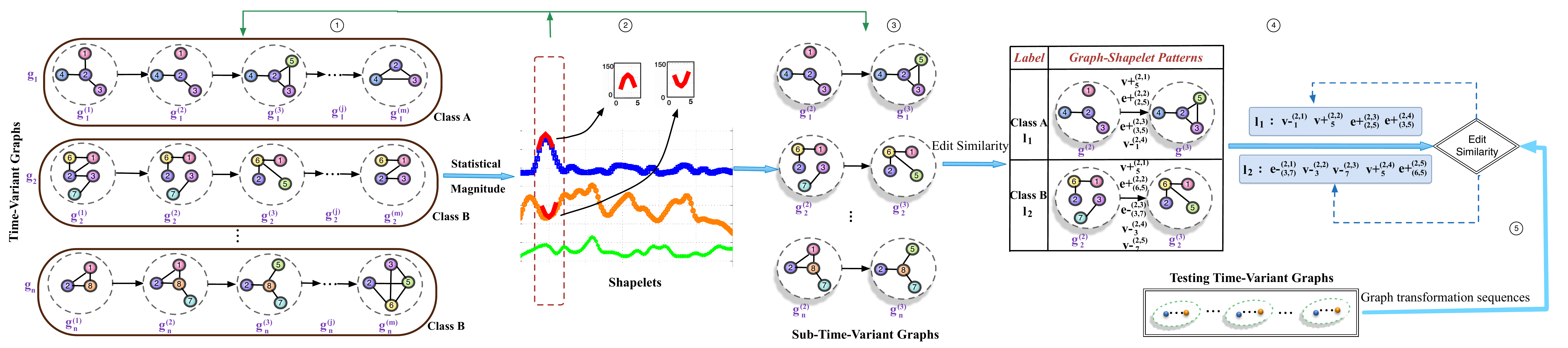}
  \caption{A concept view of the proposed time-variant graph classification framework. We first explore univariate time series from time-variant graphs via a sample kernel method in each graph (step $\textcircled{1}$, Section~\ref{gs}). Then, we find shapelet patterns from the time series by using shapelet pattern mining (step $\textcircled{2}$, Section~\ref{fg}). Next, we locate the sub-time-variant graphs that match the shapelet patterns from the original time-variant graphs (step $\textcircled{3}$, Section~\ref{ftp}). Note that each sub-time-variant graph corresponds to a unique graph transformation subsequence by the proposed time-variant graph representation approach. At the last step, we calculate the graph edit similarity between graph transformation subsequences and find the most discriminative transformation subsequences as graph-shapelet patterns (step $\textcircled{4}$, \ref{CGS}). Step $\textcircled{5}$ shows the process of time-variant graph prediction.}
\label{framework}
\end{figure*}

A transformation from $g_i^{(k)}$ to $g_i^{(k+1)}$ is represented by $op_{g_i}^k$, and $op_{g_i}^k=\langle op_{g_i}^{(k,1)}\cdots op_{g_i}^{(k,l_k)}\rangle$ with $l_k$ denoting the length of $op_{g_i}^k$,
is a series of transformations such as insertions and deletions of vertices and edges between neighboring graphs, \textit{e.g.}, $v+, e+, e-, v-$.

\begin{mydef}{(\emph{Graph Transformation Sequence and Graph Transformation Subsequence})}
Given a time-variant graph $g_i = \langle g_i^{(1)} \cdots g_i^{(m)}\rangle$, the corresponding  graph transformation sequence can be denoted as $ts(g_i) = \langle op_{g_i}^{1}\cdots op_{g_i}^{(m-1)}\rangle,$ with $ts(g_i^{\prime}) = \langle op_{g_i^{\prime}}^{1}\cdots op_{g_i^{\prime}}^{(m^{\prime}-1)}\rangle$ denoting a subsequence of $ts(g_i)$. $ts(g_i^{\prime}) \subseteq ts(g_i), m^{\prime}\leq m$, iff $\forall op_{g_i^{\prime}}^{k^{\prime}} \in ts(g_i^{\prime}), 1 \leq k^{\prime} \leq m^{\prime}-1, \exists op_{g_i}^{k} \in ts(g_i), 1 \leq k\leq m-1; op_{g_i^{\prime}}^{k^{\prime}} = op_{g_i}^{k}$ holds.
\end{mydef}

We can use time series subsequences to locate sub-time-variant graphs which are potential graph-shapelets. To predict a new testing time-variant graph, we can also convert it to a transformation sequence to compare the distances of time-variant graphs by calculating the distance of their corresponding transformation sequences.

In Fig. \ref{fig:example}, for a time-variant graph $g_i$, when its individual graph $g_i^{(1)}$ changes to $g_i^{(2)}$, the corresponding transformation sequence is $op_{g_i}^1 = \langle e-_{(1,2)}^{(1,1)} v-_1^{(1,2)}\rangle$, where superscripts $(1,1)$ and $(1,2)$ represent the first and second graph operations changing from graph $g_i^{(1)}$ to $g_i^{(2)}$. $e-$ and $v-$ represent the removal of an edge and a vertex. The order of the operations subscripts $(1,2)$ and $1$ represent that we remove an edge from the vertex of ID 1 to ID 2, and remove a vertex of ID 1. Similarly, $op_{g_i}^2 = \langle v+_5^{(2,1)} e+_{(2,5)}^{(2,2)} e+_{(3,5)}^{(2,3)}\rangle$.

\emph{Example \hypertarget{1}{1}:} A sub-time-variant graph $\langle g_i^{(1)} g_i^{(2)} g_i^{(3)}\rangle$ generates a graph transformation sequence $\langle e-^{(1,1)}_{(1,2)} v-^{(1,2)}_1 v+^{(2,1)}_5 e+^{(2,2)}_{(2,5)} e+^{(2,3)}_{(3,5)}\rangle$ as shown in Fig.~\ref{fig:example}, so we can also generate its transformation subsequence $\langle v-^{(1,2)}_1 e+^{(2,2)}_{(2,5)} e+^{(2,3)}_{(3,5)} \rangle$ with $e+^{(2,3)}_{(3,5)}$ denoting the  edge adding operation between vertices 3 and 5.

\begin{mydef}{(\emph{Edit Similarity})}\label{def:es}
For two graph transformation sequences $ts(g_i)$ and $ts(g_j)$, we use $Sim(ts(g_i), ts(g_j))$ to denote the Edit Similarity for measuring their similarity. $ts(g_i) = \langle op_{g_i}^{1}\cdots op_{g_i}^{(m-1)}\rangle, 1 \leq k \leq m-1$, with $op_{g_i}^k=\langle op_{g_i}^{(k,1)}\cdots op_{g_i}^{(k,l_k)}\rangle$ with $l_k$ denoting the length of $op_{g_i}^k$. $Sim(ts(g_i), ts(g_j))= \sum\limits_{k,v} \sum\limits_{p=1}^{l_k} \sum\limits_{q=1}^{l_v} \hbar_{(i,j)}^{(k,p),(v,q)}$, where $ \hbar_{(i,j)}^{(k,p),(v,q)}= I(op_{g_i}^{(k,p)} = op_{g_j}^{(v,q)})$, $I(\cdot)$ is 1 if the condition inside is true; otherwise, 0.
\end{mydef}\label{distanceofsubsequences}

\begin{table}[!b]
\centering
\addtolength{\tabcolsep}{+8pt}
\renewcommand{\arraystretch}{1.1}
\caption{Symbols and notations}
\label{tab:notations}
\begin{tabular}{|c|l|}
  \hline
  \textbf{Symbols} & \textbf{Descriptions} \\
  \hline
  \emph{T} & time series \\
  \hline
  \emph{ts} & transformation sequence \\
  \hline
  \emph{op} & transformation operation \\
  \hline
  $l_k$ & the length of $op_{g_i}^k$ \\
  \hline
  $m, m^{\prime}$ & \tabincell{l}{the length of time-variant graph $g_i$\\ and sub-time-variant graph $g_i^{\prime}$}\\
  \hline
  \emph{A,B} & class label\\
  \hline
  $\mathcal{V},\emph{E}$ & vertex and edge sets\\
  \hline
  $\mathcal{V}_{v+}$, $\mathcal{V}_{v-}$ & a set of vertex insertions and deletions\\
  \hline
  ${E}_{e+}$, ${E}_{e-}$ & a set of edge insertions and deletions\\
  \hline
  \tabincell{c}{vertex insertion\\ $v+^{(j,k)}_{u}$} & \tabincell{l}{insert a vertex of ID $u$ into\\ $g_i^{(j)}$ which leads to $g_i^{(j+1)}$}  \\
  \hline
  \tabincell{c}{vertex deletion\\ $v-^{(j,k)}_{u}$} & \tabincell{l}{delete an isolated vertex of ID $u$ in\\ $g_i^{(j)}$ which leads to $g_i^{(j+1)}$}\\
  \hline
  \tabincell{c}{edge insertion\\ $e+^{(j,k)}_{(u_1,u_2)}$} & \tabincell{l}{insert an edge between two vertices\\ $u_1$ and $u_2$ into $g_i^{(j)}$ and obtain $g_i^{(j+1)}$}\\
  \hline
  \tabincell{c}{edge deletion\\ $e-^{(j,k)}_{u}$} & \tabincell{l}{delete an edge between two vertices\\ $u_1$ and $u_2$ in $g_i^{(j)}$ to obtain  $g_i^{(j+1)}$}  \\
  \hline
\end{tabular}
\begin{tablenotes}
\item[] ~~~~~$*$:~$k$ is the index of operation in a transformation sequence.
\end{tablenotes}
\end{table}

\emph{Example \hypertarget{2}{2}:} In Fig.~\ref{fig:example}, the graph transformation sequence for the graph-shapelets of class A is $\langle v+^{(2,1)}_5 e+^{(2,2)}_{(2,5)} e+^{(2,3)}_{(3,5)}\rangle$, denoted by $ts(g_{i})$. The graph transformation sequence of the corresponding sub-time-variant graph in the testing time-variant graph is $\langle v+^{(2,1)}_5 e+^{(2,2)}_{(1,5)} e+^{(2,3)}_{(2,5)} e+^{(2,4)}_{(2,8)}\rangle$, denoted by $ts(g_{t})$, $Sim(ts(g_{i}), ts(g_{t})) = 2$ because of the same sequential operations. In our method, we calculate the graph edit distance \cite{sanfeliu1983distance} by using graph edit similarity.

When comparing two operations, we ignore their superscripts. For example, $e+^{(2,3)}_{(2,5)} = e+^{(2,2)}_{(2,5)}$. We use sliding windows to extract time series subsequences.

\begin{mydef}{(\emph{Graph-Shapelet Pattern})}
From a given time-variant graph data set $\mathcal{G}$, $\mathcal{G}$ can be categorized into two classes: outbreak and non-outbreak. The graph-shapelet pattern is a graph transformation subsequence \emph{(Definition 4)} which can maximally identify an outbreak time-variant graph from a non-outbreak time-variant graph by maximizing the edit similarity \emph{Definition 5}.
\end{mydef}

\begin{figure}[t]
  \centering
  \includegraphics[width=0.48\textwidth]{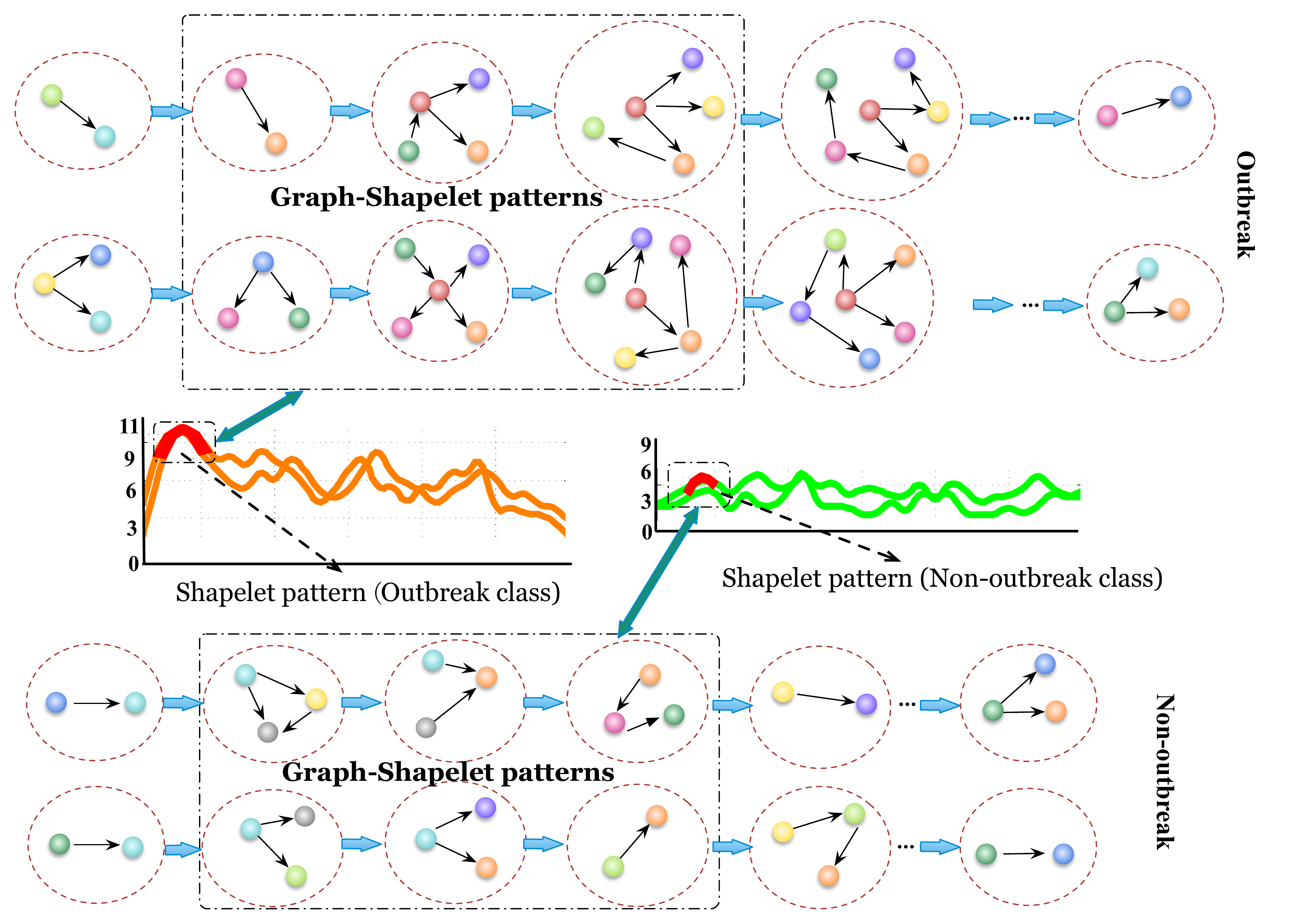}
  \caption{An illustration of graph-shapelet patterns. Graph-shapelet patterns are compact and discriminative graph transformation subsequences that describe the graph transformation patterns shared in the same class of time-variant graphs, \textit{e.g.£¬} two time-variant graphs (the two entire rows above) with the outbreak labels. When we explore a univariate time series from a time-variant graph, we can see that the location of a \emph{graph-shapelet pattern} is consistent with that of a \emph{shapelet} in time series (detailed in Section~\ref{gs}). In this case, graph-shapelet patterns can be used for time-variant graph prediction such as dynamic graph outbreak prediction.}
  \label{example_intro}
\end{figure}

Graph-shapelet patterns can be taken as a graphical extension of the traditional shapelets used in time series classification. From the time series classification viewpoint, our study extends the univariate and multi-variant time series data classification into time-variant graph classification where at each time stamp we obtain a graph instead of a single variable or a vector. As shown in Fig.~\ref{example_intro}, graph-shapelet patterns are compact and discriminative graph transformation subsequences extracted from a sequence of graph data that describes graph transformation patterns. The graph-shapelet patterns, although they looks similar to the frequent graph subsequences proposed in graph sequence mining ~\cite{Inokuchi:ICDM08}, are discriminative graph subsequences that can differentiate two classes of graph sequences with high accuracy. In contrast, frequent graph subsequences do not contain the class label information and thus are not helpful for time-variant graph classification.

When converting time-variant graphs to transformation sequences, we use the following strategies to represent the time-variant graphs and make the transformation sequences between two successive graphs unique.

\emph{Strategy 1: (The Admissibility).} When transforming two successive graphs to a transformation sequence, we fix the transformation order is fixed to make it unique, such that an insertions has precedence over a deletion, vertex insertion has precedence over edge insertion, while edge deletion has precedence over vertex deletion. Consequently, the transformation strategy between two successive graphs $g^{(j)}, g^{(i)}$ is as follows:

\begin{small}
\begin{align*}
&v+^{(j,k)}_{u_1} \prec e+^{(j,k^{\prime})}_{(u_1,u_2)},e+^{(j,k)}_{(u_1,u_2)} \prec e-^{(j,k^{\prime})}_{(u_1,u_2)},e-^{(j,k)}_{(u_1,u_2)} \prec v-^{(j,k^{\prime})}_{u_1}
\end{align*}
\end{small}
where $k < k^{\prime}$. The definitions of $v+^{(j,k)}_u, v-^{(j,k)}_u, e+_{(u_1,u_2)}^{(j,k)}, e-_{(u_1,u_2)}^{(j,k)}$ are illustrated in Table~\ref{tab:admissibility}.

In terms of the same operation with different IDs, we use a lexicographic order to generate a unique sequence, \emph{e.g.}, $v+_1^{(1,1)} \prec v+_2^{(1,2)}$.

\emph{Strategy 2: (The Representation).} Since all operations between two successive graphs can be enumerated as edge insertion, edge deletion, vertex insertion, and vertex deletion, results in any time-variant graph can be represented by the four operations in Table~\ref{tab:admissibility} with a given initial $g_i^{(1)}$ \cite{Inokuchi:ICDM08}.

Table~\ref{tab:notations} summarizes major notations used in the paper.

\section{The Proposed Method}
\label{tpm}

\subsection{The Overall Framework}
Fig.~\ref{framework} shows the framework of our method which first explore a univariate time series from each time-variant graph via a simple kernel method (Section~\ref{gs}). We then calculate shapelets from the time series using shapelets mining (Section~\ref{fg}). After that, sub-time-variant graphs that match the shapelets from the original time-variant graphs are located. Each sub-time-variant graph corresponds to a unique graph transformation subsequence by applying the Admissibility strategy and the Representation strategy. In the last step, we calculate the graph edit similarity between graph transformation subsequences and find the most discriminative transformation subsequences as graph-shapelets (Section~\ref{ftp}).

\subsection{Graph-Shapelet Pattern Location}
\label{gs}

\begin{algorithm}[!t]
\caption{{SubTVG: Sub-time-variant graph candidates}}\label{alg:alg1}
\begin{algorithmic}[1]
\REQUIRE ~~\\
    $\mathcal{G}$: A set of time-variant graphs;\\
    $l$: Length of graph-shapelet patterns;\\
\ENSURE ~~\\
    $\mathcal{G}^{\prime}$: Graph subsequence candidates;\\

\STATE $D \leftarrow$ Apply $\mathcal{G}$ to obtain corresponding time series $D$;
\STATE $cSet\leftarrow\emptyset$; // Time series candidates with length $l$ \\
//Generate time series candidate with length $l$
\FOR {$t$ \textbf{in }$D$}
    \STATE $S^l_t \leftarrow$ Extract time series subsequence with length $l$ from $t$
    \STATE $cSet \leftarrow cSet \bigcup \mathcal{S}_t^l$;
\ENDFOR
\STATE Shapelets $\leftarrow$ Apply $cSet$ to $D$ based on shapelet mining\cite{Ye:KDD09}.
\STATE Record the Shapelet locations in each time series
\FOR {\textbf{each} $g_i$ \textbf{in} $\mathcal{G}$}
    \STATE $g_i^{\prime} \leftarrow$ Apply recoded Shaplelet locations to the related sub-time-variant graphs in $g_i$;
    \STATE $\mathcal{G}^{\prime} =  \mathcal{G}^{\prime} \bigcup g_i^{\prime}$
\ENDFOR
\RETURN {$\mathcal{G}^{\prime}$}

\end{algorithmic}
\end{algorithm}

Although many methods exist to find shapelets for time series classification, it is very challenging to restore sub-time-variant graphs from the time series shapelets. To address this challenge, we record the locations of time series segments that have the minimum Euclidean distance, while introducing a transformation of time-variant graphs to the time series.

We first design a kernel method to explore time-series information $t_i=\{t_i^{(1)},\cdots,t_i^{(m)}\}$ from a time-variant graph $g_i$, $K_{t_i}^{(j)} = N_n(g_i^{(j)}) + N_e(g_i^{(j)})$, where $N_n(g_i^{(j)})$ and $N_e(g_i^{(j)})$ denotes the statistical information of the vertices and edges in an individual graph $g_i^{(j)}$ at time $j$ with $1 \le j \le m$. After that,
efficient shapelet pattern mining algorithms can be applied to quickly locate graph changes in time-variant graphs. The reason why we use the graph statistic method is that we can find the location of graph-shapelets and return to graphs after discovering graph-shapelets. By using this method, we can speed up the search time and save the graph information since we only use a segment of a time-variant graph.

\emph{Example \hypertarget{3}{3}:} In Fig.~\ref{fig:example}, the statistics of vertices and edges in graphs $g_i^{(1)}, g_i^{(2)}, \cdots, g_i^{(m)}$ are $7, 5, \cdots , 6$ respectively. Thus, a corresponding time series $\{7, 5, \cdots, 6\}$ can be explored. Note that, we use the sum of the statistical information of the vertices and edges (\emph{i.e.}, $K_{t_i}^{(j)}$) when transforming a graph in a time-variant graph to a numerical value in the corresponding time-series

\subsection{Graph-Shapelet Pattern Candidates}
\label{fg}

All sub-time-variant graphs are potential graph-shapelet candidates. The maximum similarity of a candidate to all training sequences can be used for graph prediction by ranking the similarity of a candidate \textit{w.r.t.} a testing time-variant graph. To find graph-shapelet patterns, the algorithm needs to generate potential sub-time-variant graphs as candidates and to locate the one, which has maximum similarity \textit{w.r.t.} other subsequences from the candidate sub-time-variant graphs.

Algorithm~\ref{alg:alg1} shows the detailed procedures for finding sub-time-variant graph candidates from time-variant graphs. Given a time-variant graph database $\mathcal{G}$ with a user-defined graph-shapelet length $l$. In Algorithm \ref{alg:alg1}, line 1 converts time-variant graph set $\mathcal{G}$ to a time series \emph{D} based on the number of vertices and edges in each graph. Specifically, it calculates $K_{t_i}^{(j)}$ based on $N_n(g_i^{(j)})$ and $N_e(g_i^{(j)})$ in $\mathcal{G}$. Thus, we can obtain the corresponding time series \emph{D}. Then the algorithm generates all candidates within a sliding window (lines 3-6). Line 4 is to divides an input time series into a sequence of discrete segments by using a sliding window. The sliding window algorithm works by anchoring the left point of a potential segment at the first data point of a time series, then attempts to approximate the data to the right with increasing longer segments of length \emph{l}. Line 7 mines shapelets from time series \emph{D}. The most straightforward way for finding the shapelet is using the brute-force method. Given the time series \emph{D} and the user-defined maximum and minimum lengths of the shapelet, we first generate all of the segments with length \emph{l}. Then the algorithm checks how well each segment can separate \emph{D} in class \emph{A} (outbreak) and class \emph{B} (non-outbreak). For each shapelet candidate, the algorithm calculates the information gain \cite{Ye:KDD09} achieved if using that candidate to separate the data. The algorithm returns the candidates with the highest information gain as shapelets. The reason we use an exhaustion method for finding shapelets is that the transformed time series from the time-variant graph is short so that the search space is low. Thus, the simple yet efficient brute-force shapelet search algorithm is applicable in this case. Since the shapelet mining method can return the locations of shapelets, we find the same locations from the time-variant graph to discover the graph-shapelet (lines 9-12).

\begin{algorithm}[!t]
\caption{{GraphSequenceCompiler}}\label{alg:alg2}
\begin{algorithmic}[1]
\REQUIRE ~~\\
    $\mathcal{G}^{\prime}$: time-variant graph candidates;
\ENSURE ~~\\
    $ts$: graph transformation sequences;

\STATE $ts\leftarrow \emptyset$;
\FOR {\textbf{each} $g_i \in \mathcal{G}^{\prime}$}
    \FOR {\textbf{each} $g_i^{(j)} \in g_i$}
        \STATE Represent $g_i^{(j)}$ to $\mathcal{V}_{v+}, \mathcal{V}_{v-}, E_{e+}, E_{e-}$ follow \emph{Strategies 1, 2};
        \STATE $k \leftarrow 1$
        \FOR {\textbf{each} $v$ in $\mathcal{V}_{v+}$}
            \STATE $ts \leftarrow ts \bigcup {v+}^{(j,k++)}_{id(v)}$; ~~// Get the insertion vertices set
        \ENDFOR
        \FOR {\textbf{each} $(v, v^{\prime})$ in $E_{e+}$}
            \STATE $ts \leftarrow ts \bigcup {e+}^{(j,k++)}_{(id(v),id(v^{\prime}))}$;~~// Get the insertion edges set
        \ENDFOR
        \FOR {\textbf{each} $(v, v^{\prime})$ in $E_{e-}$}
            \STATE $ts \leftarrow ts \bigcup {e-}^{(j,k++)}_{(id(v),id(v^{\prime}))}$;~~// Get the deletion edges set
        \ENDFOR
        \FOR {\textbf{each} $v$ in $\mathcal{V}_{v-}$}
            \STATE $ts \leftarrow ts \bigcup {v-}^{(j,k++)}_{id(v)}$;~~// Get the deletion vertices set
        \ENDFOR
    \ENDFOR
\ENDFOR
\RETURN $ts$
\end{algorithmic}
\end{algorithm}

\begin{algorithm}[!t]
\caption{{Graph-shapelet Pattern Classifier}}\label{alg:alg4}
\begin{algorithmic}[1]
\REQUIRE ~~\\
    $\mathcal{G}$: a set of $n$ time-variant graphs;\\
    $l$: Length of the graph-shapelets; \\

\ENSURE ~~\\
     $L_t$: The label of a test time-variant graph $g_t$; \\
// \textbf{Training Phase:}
\STATE $graph\_shapelt \leftarrow \emptyset$;\\
\STATE $candidates \leftarrow$ SubTVG$(\mathcal{G}, l)$;   // Algorithm 1
\STATE $max\_sim \leftarrow 0$;
    \STATE $ts \leftarrow$ GraphSequenceCompiler($candidates$);   // Algorithm 2
    \FOR {each $ts_i \in ts$}
        \STATE $d_i \leftarrow \sum_j Sim(ts_i, ts_j), ts_j \in ts (i \neq j)$;\\
        \IF {$d_i > max\_sim$}
            \STATE $max\_sim \leftarrow d_i$;\\
            \STATE $graph\_shapelets \leftarrow ts_i$;\\
        \ENDIF
    \ENDFOR \\

// \textbf{Test Phase:}
    \STATE $ts(g_t) \leftarrow$ GraphSequenceCompiler($g_t$);   // Algorithm 2
    \FOR {each $graph\_shapelet$ in $graph\_shapelets$}
        \STATE calculate $Sim(graph\_shapelet, g_t)$;\\

\ENDFOR
\STATE $L_t$ is the label with maximum similarity;\\
\RETURN $L_t$
\end{algorithmic}
\end{algorithm}

\subsection{Finding Graph-Shapelet Patterns}
\label{ftp}
Graph-shapelet patterns can be found from the generated sub-time-variant graph candidates. To calculate graph-shapelet patterns from sub-time-variant graphs, time-variant graphs are converted to transformation sequences so that edit similarity can be used as the measure. The transformation sequences between two graphs in a given time-variant graph can be represented by a series of operations, \textit{e.g.}, vertex and edge insertions and deletions based on \emph{Strategy 1} and \emph{Strategy 2}. According to the time-variant graph representation, an algorithm to compile $\mathcal{G} = \{g_i|g_i = \langle g_i^{(1)} \cdots g_i^{(m)}\}$ to a set of transformation sequences $ts(\mathcal{G}) = \{ts(g_i)|g_i \in \mathcal{G}\}$ is summarized in Algorithm~\ref{alg:alg2}.

To obtain the graph transformation sequence \emph{ts} in Algorithm \ref{alg:alg2}, we first represent each two successive graphs in the time-variant graph $\mathcal{G}$ as transformation sequences. \emph{Strategy 2} indicates that there are four operations (vertex insertion, vertex deletion, edge insertion and edge deletion) between two successive graphs. \emph{Strategy 1} limits the order of the operations to ensure a unique transformation sequence. For example, for the graph sequence $<g_i^{(2)} g_i^{(3)}>$ shown in Fig.~\ref{fig:example}, the transformation sequence can be $<v+_5^{(2,1)}, e+_{(2,5)}^{(2,2)},e+_{(3,5)}^{(3,3)}>$ or $< e+_{(2,5)}^{(2,2)}, v+_5^{(2,1)}, e+_{(3,5)}^{(3,3)}>$ if we do not follow \emph{Strategy 1}. Therefore, the procedure basically follows \emph{Strategies 1 and 2} to convert graph sequences (line 4). Then, the vertex and edge insertions are appended to the transformation sequences, as shown on lines 6-8 and 9-11 respectively. After that, vertex and edge deletions are carried out on lines 12-14 and 15-17 respectively, with the changes appended to the transformation sequences.

\emph{Example \hypertarget{4}{4}:} In the graph sequence $\langle g_i^{(2)} g_i^{(3)}\rangle$ given in Fig.~\ref{fig:example}, the insertion of vertex ID 5 and edge insertions $(2,5)$ and $(3,5)$ from $g_i^{(2)}$ to $g_i^{(3)}$, result in the transformation sequences, \emph{i.e.}, $v+^{(2,1)}_{5}, e+^{(2,2)}_{(2,5)}, e+^{(2,3)}_{(3,5)}$.

\subsection{Classification with Graph-Shapelet Patterns}\label{CGS}

Fig.~\ref{framework} gives an example of finding two graph-shapelet patterns for a binary classification problem, and the algorithm details are summarized in Algorithm~\ref{alg:alg4}. First, the algorithm calls SubTVG() to generate all graph-shapelet pattern candidates (line 2) and calls function GraphSequenceCompiler() to generate graph transformation sequence for each generated sub-time-variant graph candidate (line 4). The similarity between each generated sub-time-variant graph and all time-variant graphs in each class (line 6) is calculated, and the sub-time-variant graphs with maximum similarity are selected as the graph-shapelet patterns, and the transformation sequences used as classification features (lines 7-10).

Whenever a test time-variant graph arrives, the algorithm converts it to transformation sequences and computes the distance between the test transformation sequences and features selected for different classes by using Edit Similarity (Definition 5). The output label of the test graph is the same as the label of the graph-shapelet pattern, which has maximum similarity to the test time-variant graphs (lines 12-17).

The proposed method is inspired by shapelet features in a time series. Since the difference between time series and time-variant graphs is that the time-variant graph is constituted by a graph at each time point instead of a numerical value, it is reasonable that we convert the given time-variant graphs to time series. As we find graph-shapelets based on the shapelets from converted time series, the graph-shapelet patterns are a discriminative segment in the time-variant graph. Consequently, we can only consider the graph-shapelet instead of the whole graph sequence.

\subsection{Time Complexity Analysis}
The proposed graph-shapelet pattern based time-variant graph classification algorithm consists of (1) converting time-variant graphs to graph transformation sequences, and (2) finding frequent transformation subsequences as graph-shapelet patterns for classification. We assume $m$ is the length of the query time-variant graphs, $l$ is the length of the graph-shapelet pattern, $r$ is the average length of the transformation sequences between two graphs, and $n$ is the size of the time-variant graph data set. The algorithm first transfers time-variant graphs to time series, which scans each graph (length is $m$) over all the time-variant graph samples (length is $n$). Then, the algorithm extracts segments with length $l$ (the same length with graph-shapelet patterns) to find shapelets from the time series. Therefore, the time complexity for graph-shapelet mining is $O(ml*n)$. After finding the graph-shapelets, the algorithm generates the corresponding graph transformation sequences. Since the algorithm generates every possible operation in the transformation sequences one by one, the entire graph classification time complexity is $O(mrl*n)$.

\section{Experiments}
\label{experiments}
\subsection{Data sets}
\subsubsection{Synthetic Time-Variant Graph Data}

In our experiments, we first create synthetic information propagation to validate algorithm performance on data with known groundtruth properties. We use a well-known model, namely Small World~\cite{Watts:Nature98}, to generate synthetic information propagation graphs for testing and comparison. This model generates a ring of $N$ vertices where each vertex is connected to its $\alpha$ nearest neighbors in the ring ($\alpha/2$ on each side, which means $\alpha$ must be even). Then, for each edge in the graph, the target vertex with probability $p$ is rewired. The rewired edge cannot be a duplicate or self-loop. In particular, we set $\alpha=4$ and $p=0.1$ in our experiments for synthetic time-variant graph generation. After generating the first graph $g_0^{(0)}$, we simulate the propagation from $g_0^{(0)}$ to generate the other graphs in the first time-variant graph sequence $g_0$ using the following method. First, we randomly select a root vertex \emph{r} with a non-zero out-degree from $g_0^{(0)}$. The vertex \emph{r} is then added to the initially empty list \emph{I} and all outgoing edges \emph{(r; s)} are added to the initially empty first-in-first-out (FIFO) queue, where the first element added to the queue will be the first one to be removed. We choose an edge from the candidate set each time and calculate the time delay for the edge until the time delay exceeds a given time window (we use [0,1000] as the time window). When calculating time delay of propagation, we follow a power-law distribution exponent $\beta = 1.5$, which corresponds to the power-law degree exponent of the Weibo network. Finally, we separate graph $g_0^{(1)}, \cdots, g_0^{(i)}$ with time delay 100, 200, 300, 400, 500, 600 and 700 (\emph{e.g.}, we form the first graph in a time-variant graph when the propagation time delay is 100). Thus, the average length in the synthetic data is 7 since we have 7 graphs in each time-variant graph. The whole data set is generated by repeating the above steps to generate 200 graph sequences. If the number of vertices in a graph sequence is more than 100 (a user-defined threshold based on the characteristics of the data set), we label it as an outbreak class, otherwise as a non-outbreak class. The learning task is to predict whether a time-variant graph will be an outbreak or not.

\subsubsection{Real-World Time-Variant Graph Data}

We evaluate the performance of the proposed method on two real-world data sets, including phone call graph obtained from MIT \cite{Eagle:NAS09} and information propagation graph data crawled from Sina Weibo, which is one of the biggest social media platforms of China. Akin to a hybrid of Twitter and Facebook, Weibo users can post messages, and upload pictures and videos for sharing in real-time. Users can also comment and forward posted messages

The phone call time-variant graph contains the dynamics of 75 students/faculty members in the MIT Media Laboratory, and 25 incoming students at the MIT Sloan Business School adjacent to the Media Laboratory. The experiment was designed to study community dynamics (classified as personal behavior or interpersonal interactions), by tracking a sufficient number of people using their personal mobile phones. In our experiments, a unique ID is assigned to each person participating in the communication and an edge links two persons if they communicate via phone call in a day. In this way, we obtain a daily graph $g^{(j)}$. According to the daily vertex degree (threshold is set to 5), each person is either a hub person or a non-hub person. A person with a high vertex degree is considered to be a hub person in the community. We obtain a set of weekly time-variant graph data, \emph{i.e.}, $\mathcal{G}$, and the total number of weeks (number of sequences) is 40. We randomly sample $|V| = (1\sim20)$ persons, and assign the label 1 if the weekly phone call includes a hub person (\textit{i.e.,} a hub person usually triggers interpersonal interactions), otherwise the label 0 is applied. The average length of each sequence is 7. The Weibo time-variant graph data set is obtained from a research community in Weibo by crawling 200 Weibo message propagations from a research community within one month. We sample 50 users and assign each of them a vertex ID (varying from $1$ to $50$). The propagation of the information from users follows a temporal order.
At each time point, all vertices reached by the Weibo message form a graph. For a new Weibo message, the forwarding re-forwarding propagation graphs collected in a time period form a time-variant graph (\textit{i.e.} a time-variant graph), and our learning goal is to use these time-variant graphs to predict whether the time-variant graph for a new Weibo message will break out or not. In our experiments, we obtain 200 time-variant graphs from the propagation of 200 Weibo messages at different time periods, and the average length of each sequence is 15. We define a time-variant graph as an outbreak if the total number of vertices is more than 100.

\begin{figure*}[!t]
 \centering
  \subfigure[Synthetic time-variant graph]{\includegraphics[width=0.32\textwidth]{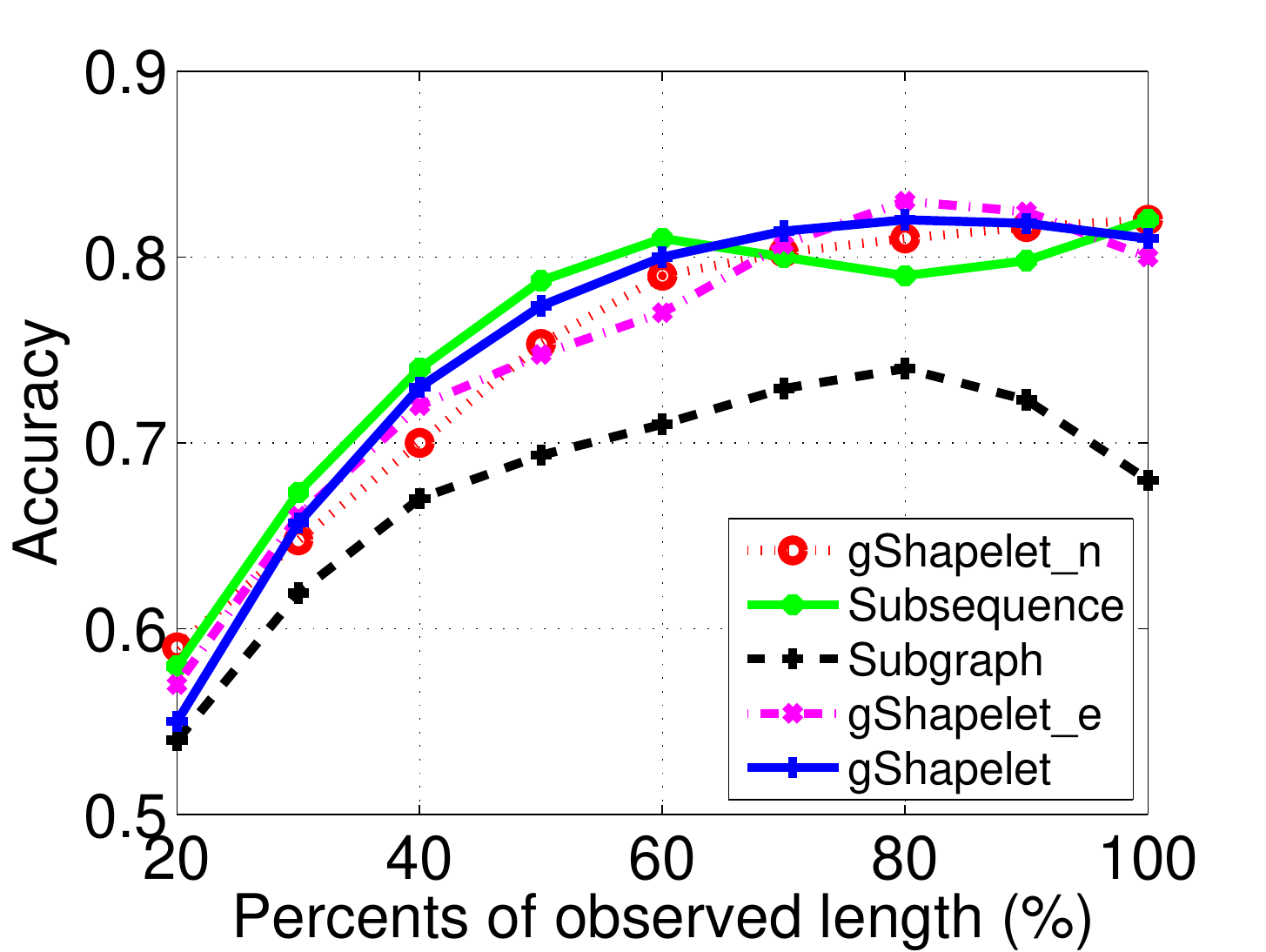}}
  \subfigure[Phone call time-variant graph]{\includegraphics[width=0.32\textwidth]{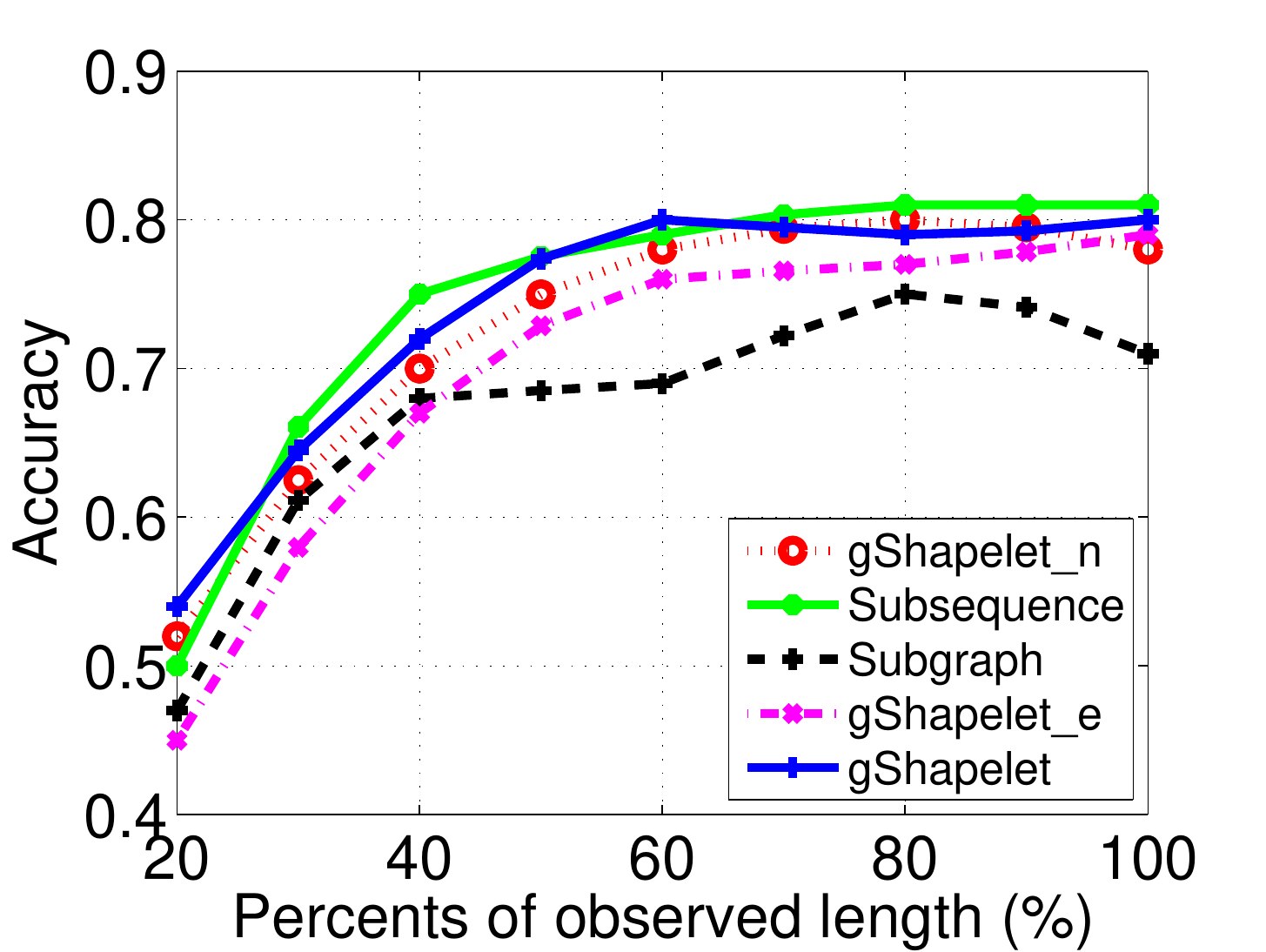}}
  \subfigure[Weibo time-variant graph]{\includegraphics[width=0.32\textwidth]{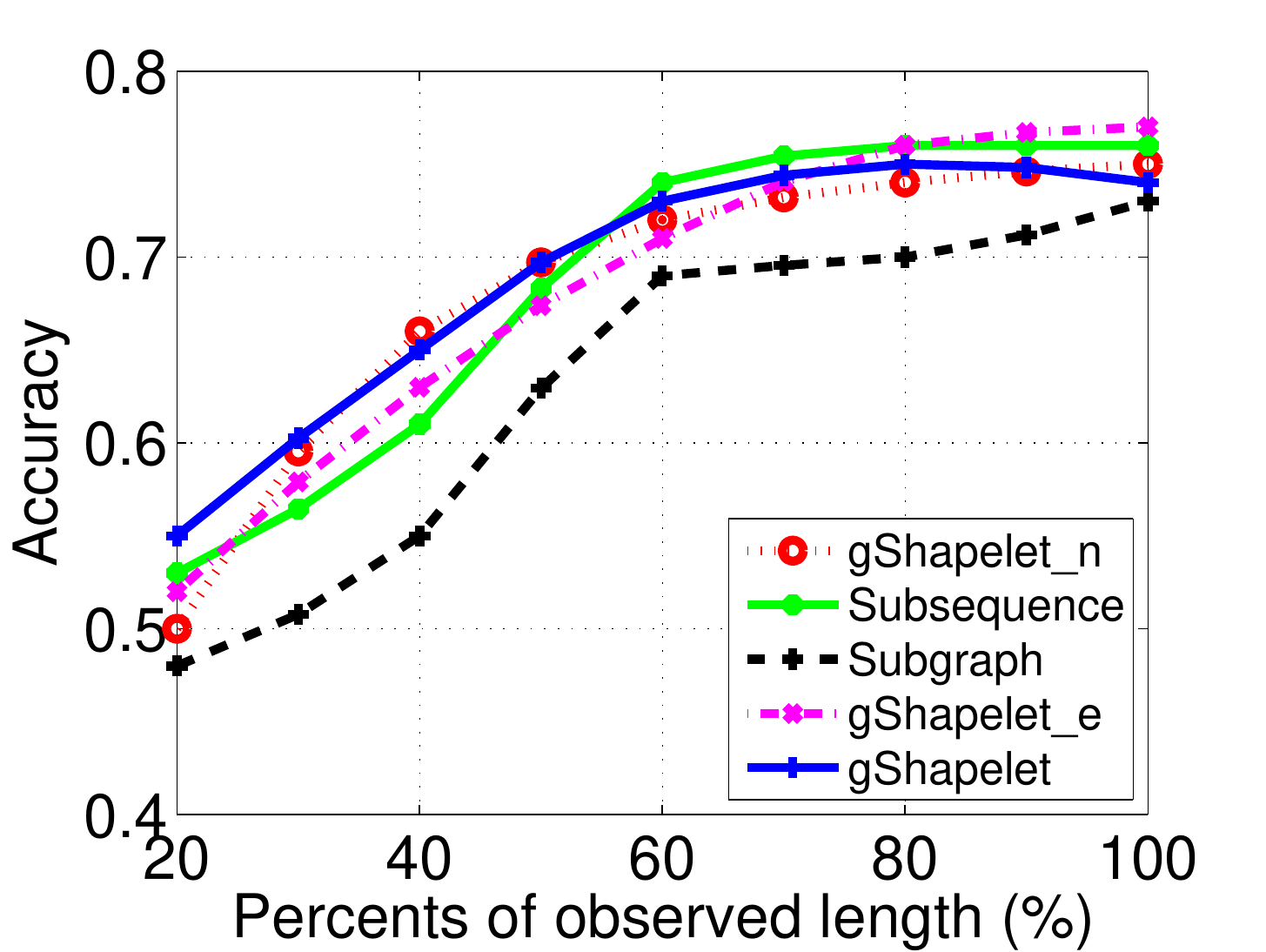}}
    \caption{Accuracy comparisons with respect to different time stages on both synthetic and real-world time-variant graph data sets.}
  \label{fig:acc}
\end{figure*}

\begin{figure*}[!t]
 \centering
  \subfigure[Synthetic time-variant graph]{\includegraphics[width=0.32\textwidth]{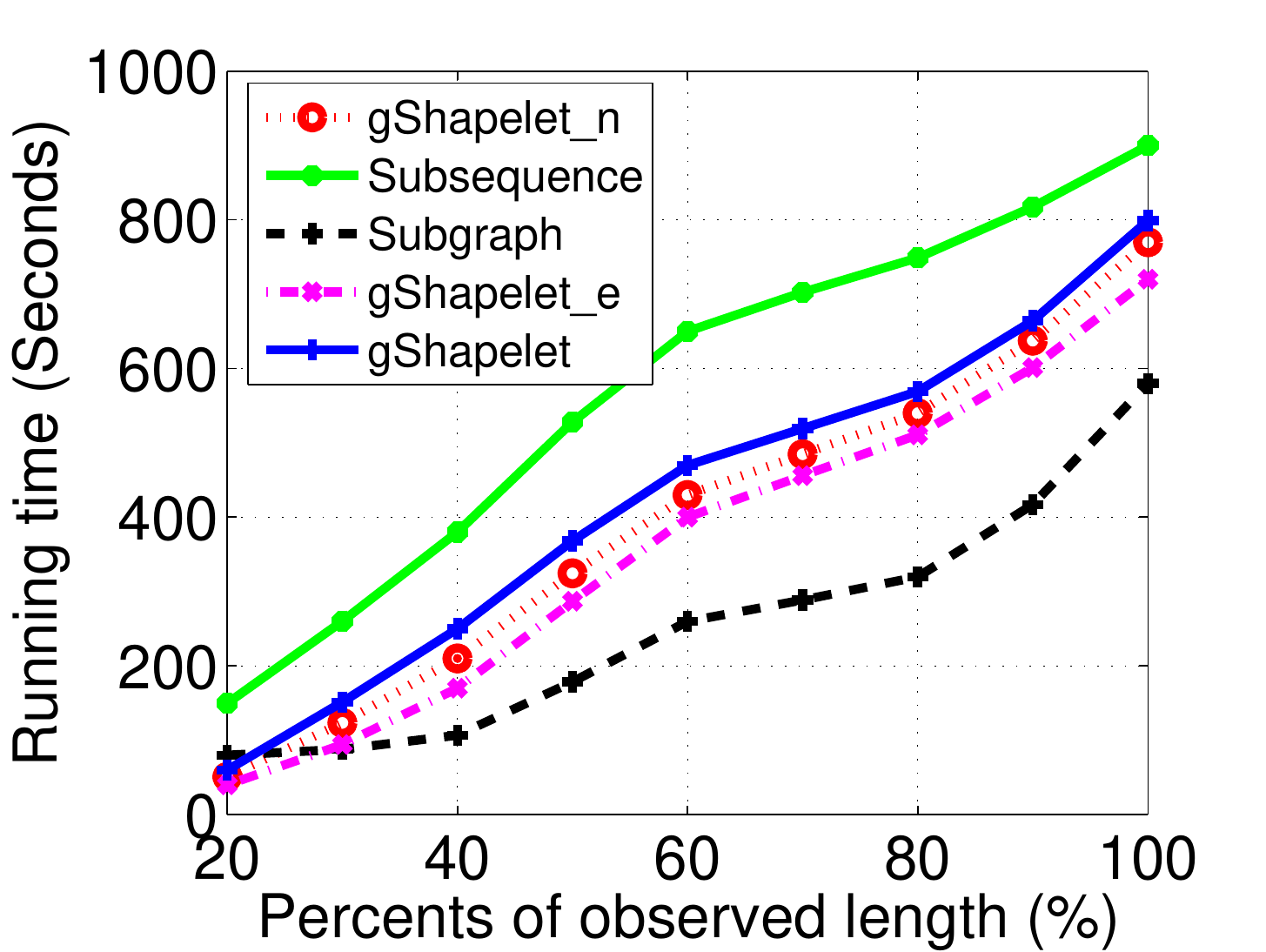}}
  \subfigure[Phone call time-variant graph]{\includegraphics[width=0.32\textwidth]{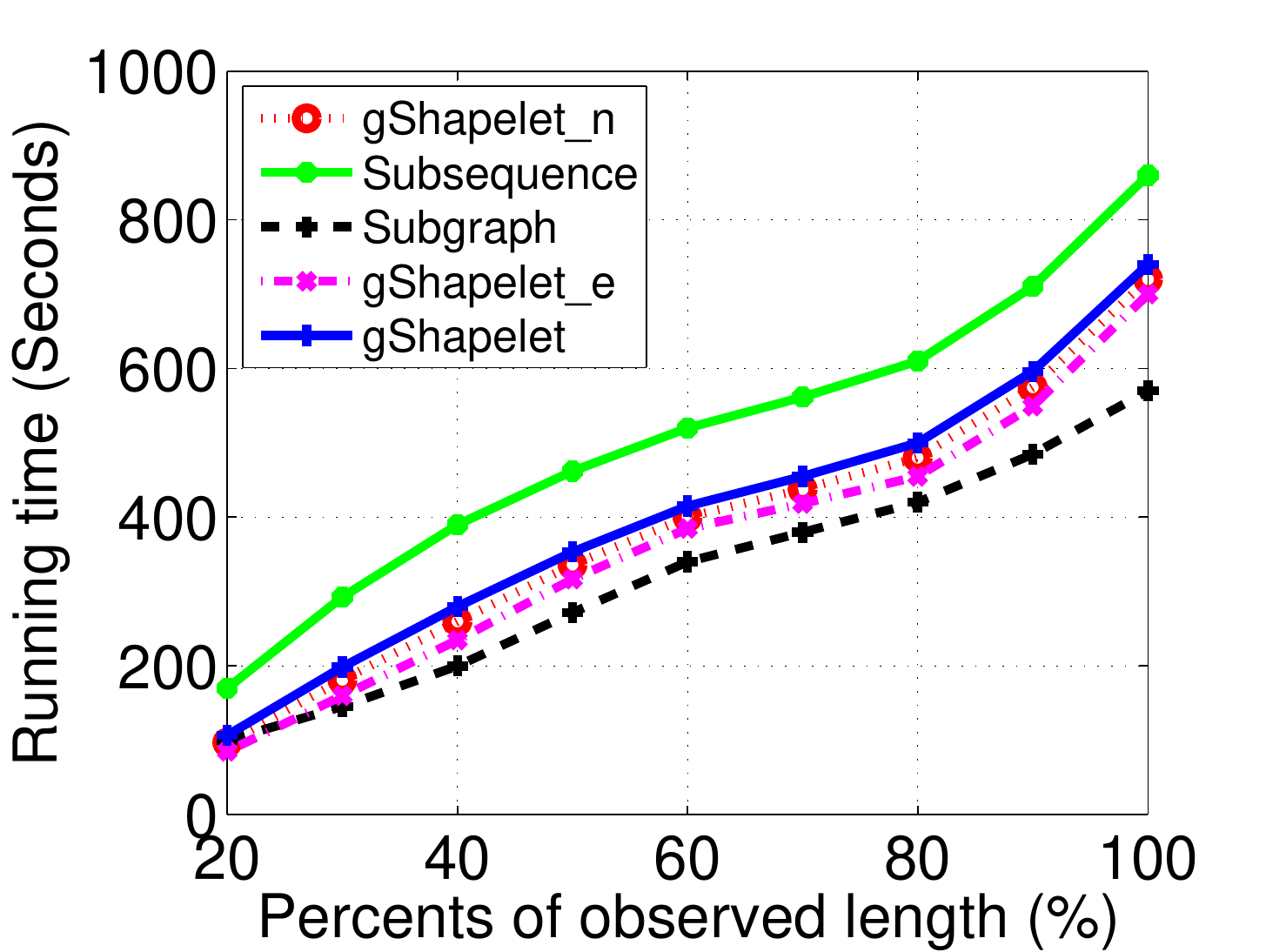}}
  \subfigure[Weibo time-variant graph]{\includegraphics[width=0.32\textwidth]{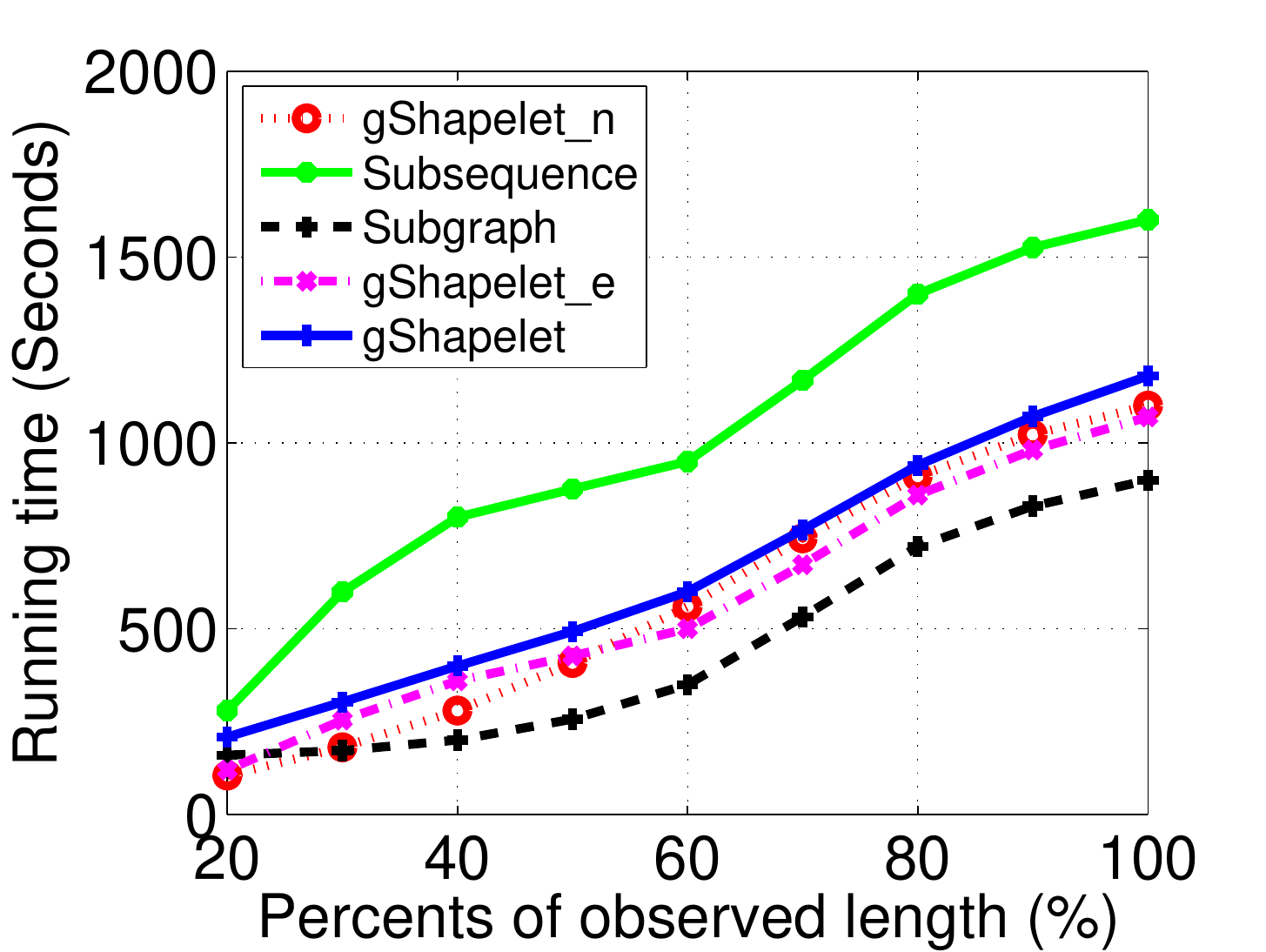}}
    \caption{The average CPU time with respect to different time stages on both synthetic and real-world time-variant graph data sets.}
  \label{fig:runningtime}
\end{figure*}

\begin{figure*}[!t]
 \centering
  \subfigure[Synthetic time-variant graph data]{\includegraphics[width=0.32\textwidth]{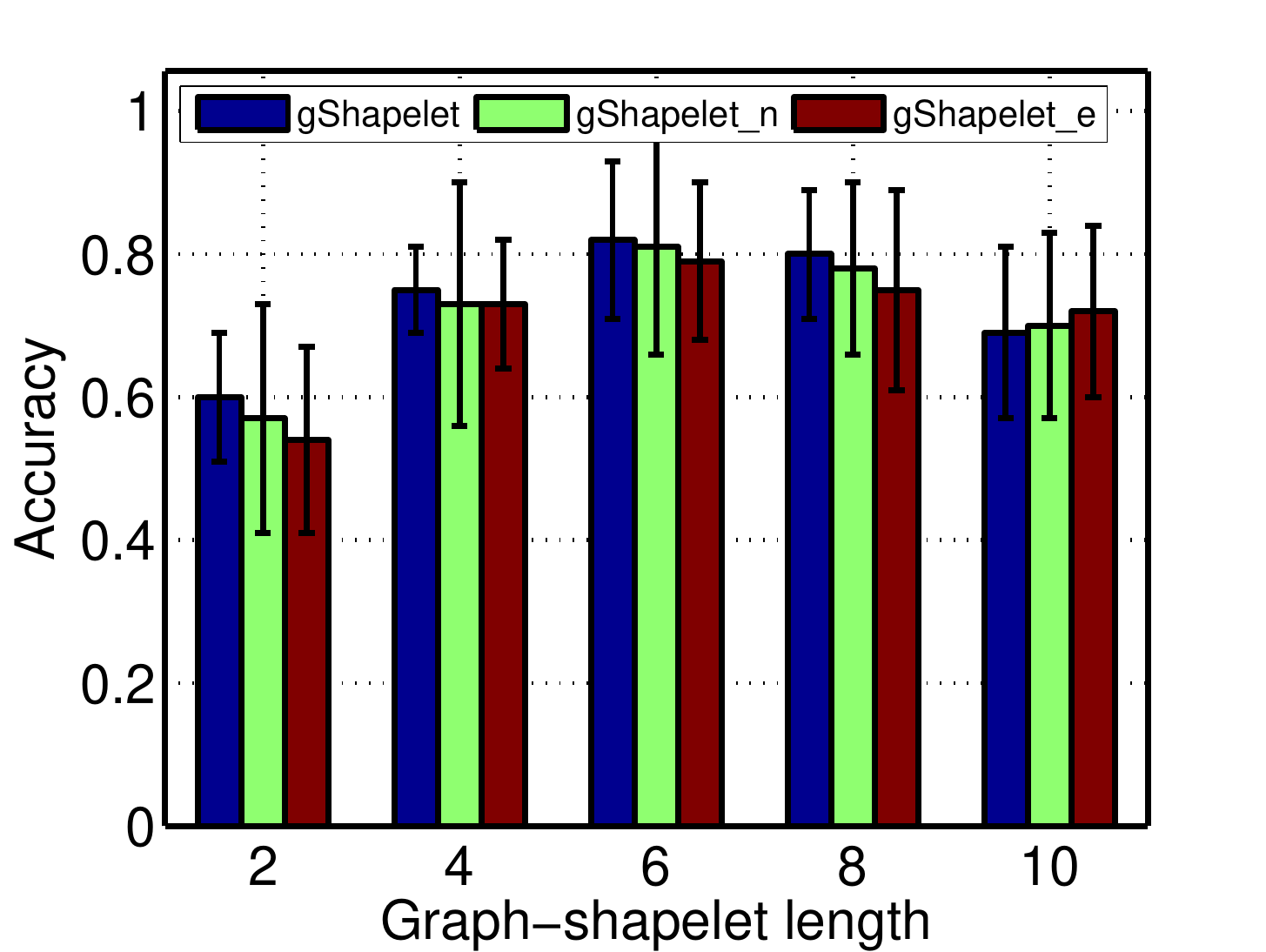}}
  \subfigure[Phone call time-variant graph data]{\includegraphics[width=0.32\textwidth]{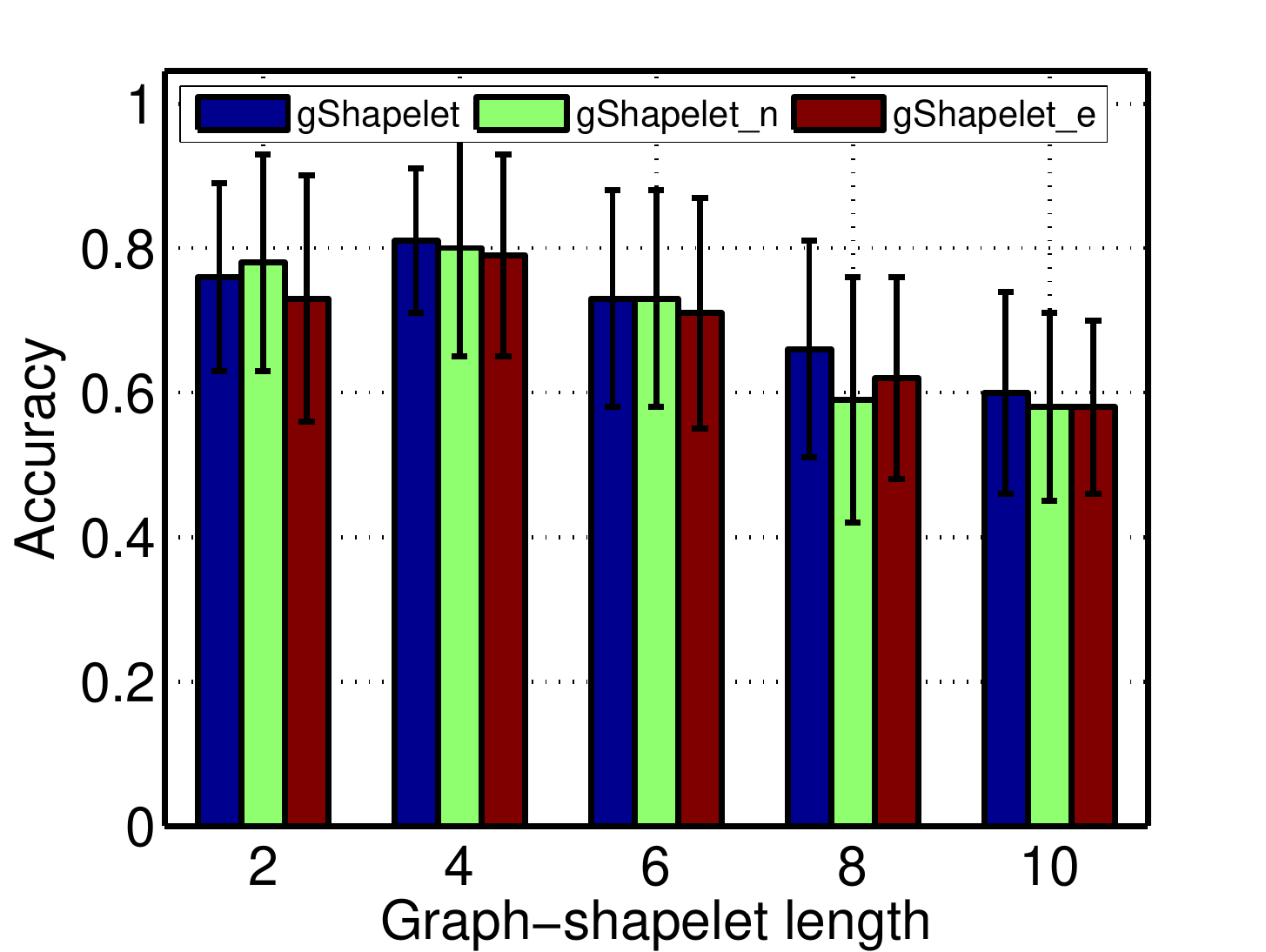}}
  \subfigure[Weibo time-variant graph data]{\includegraphics[width=0.32\textwidth]{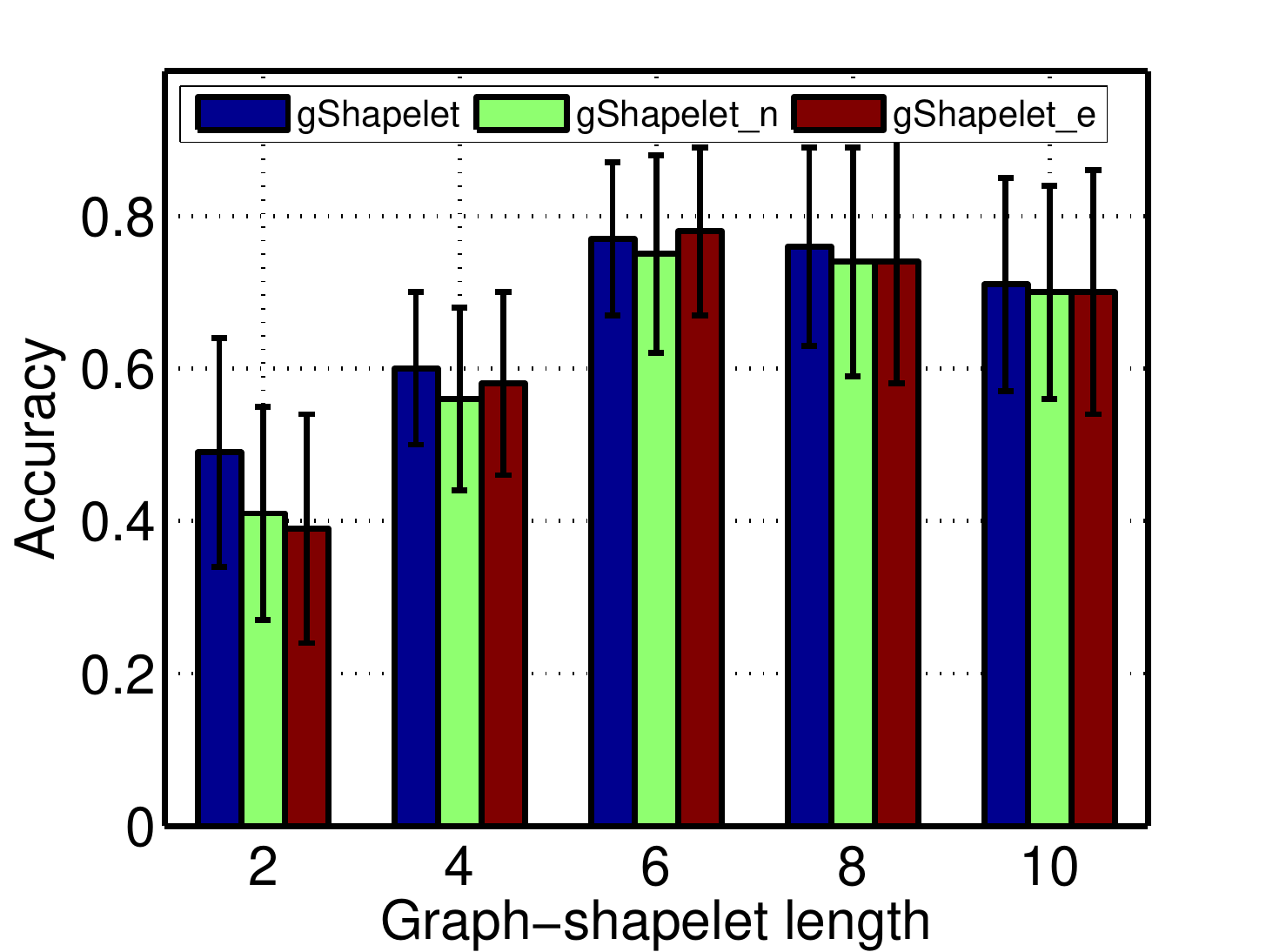}}
    \caption{Comparisons with varying graph-shapelet length on both synthetic and real-world time-variant graph data sets.}
  \label{fig:lentest}
\end{figure*}

\subsection{Experimental Settings}
\subsubsection{Baseline Approaches}
Because no existing approaches are available for time-variant graph classification, we compare the proposed graph-shapelet patterns based algorithm (gShapelet for short) with the following benchmark methods in terms of prediction accuracy and running time to evaluate classification performance.
\begin{itemize}
  \item \textbf{Frequent subsequence based method}. This type of method converts all the training time-variant graphs to graph transformation sequences and discovers frequent subsequences as features for classification. This method resembles the work in~\cite{Inokuchi:ICDM08}, the difference being that \cite{Inokuchi:ICDM08} does not consider classification problems.
  \item \textbf{Frequent subgraph based method}. Because we cannot obtain stable subgraphs from time-variant graphs, we mine frequent subgraphs from the last graph in a time-variant graph as features via gSpan~\cite{Yan:ICDM02}, a well used frequent subgraph method.
  \item \textbf{gShapelet\_n}. When a time-variant graph is transformed into a univariate time series, gShapelet\_n counts the number of vertices as statistical magnitude.
  \item \textbf{gShapelet\_e}. To transform a time-variant graph into a univariate time series, gShapelet\_e counts the number of edges as statistical magnitude.
\end{itemize}

\subsubsection{Evaluation Measures}
We randomly select 60\% time-variant graphs for training, and the remaining time-variant graphs are used for testing. When using frequent subsequence or subgraph based methods, we need to set the support threshold $\sigma$. Given a set of data $\mathcal{G} = \{g_i|g_i = \langle g_i^{(1)} \cdots g_i^{(m_i)}\rangle\}$, a support value $\sigma$ of a transformation subsequence and subgraph in graph sequence $g^{\prime}$ are defined as $$\sigma = \frac{|\{g_i|g_i \in \mathcal{G}, ts(g^{\prime}) \sqsubseteq ts(g_i)\}|}{|\mathcal{G}|}~ for~ subsequences,$$ $$\sigma = \frac{|\{g_i|g_i \in \mathcal{G}, g^{\prime} \sqsubseteq g_i\}|}{|\mathcal{G}|}~ for ~ subgraphs.$$

\subsection{Experimental Results}
The four algorithms are compared under different observation lengths of the time-variant graphs, \emph{i.e.}, 20\%, 40\%, 60\%, 80\% and 100\% in each time-variant graph. In Fig.~\ref{fig:acc} and Fig.~\ref{fig:runningtime}, we report the algorithm performance in terms of classification accuracy and running time, respectively.

\subsubsection{Effectiveness Results}
The results in Fig.~\ref{fig:acc} show that classification accuracy increases with respect to observation time, and the results tend to be stable as the time elapses. This is mainly because more vertices are observed which leads to more accurate shapelet or subsequence or subgraph patterns. Once the participant length reaches 60\%, the algorithms are likely to find the best graph-shapelet patterns, so the accuracy tends to become stable. These results imply that graph-shapelet patterns can be used for early prediction.

As we can observe from Fig. \ref{fig:acc}, the graph-shapelet based approaches have comparable accuracy with the subsequence-based method and perform much better than the subgraph-based method. This is because the subsequence-based method mines frequent features from the entire time-variant graphs. This kind of approach ensures classification accuracy, but it requires more running time. The graph-shapelet based methods first find discriminative segments from the time-variant graphs, and then find features from the segments instead of the whole time-variant graphs. This method results in a significant reduction in running time. Consequently, as with on-line and off-line methods, accuracy is traded for running time performance, especially on large-scale data sets. Our aim is to make the proposed methods more efficient under comparable accuracy. The proposed graph-shapelet methods outperform the subgraph-based method because the latter does not consider the temporal correlations during subgraph mining, and the distance between subgraph patterns may change irregularly over time.

\begin{figure}[!b]
  \centering
  \includegraphics[width=0.45\textwidth]{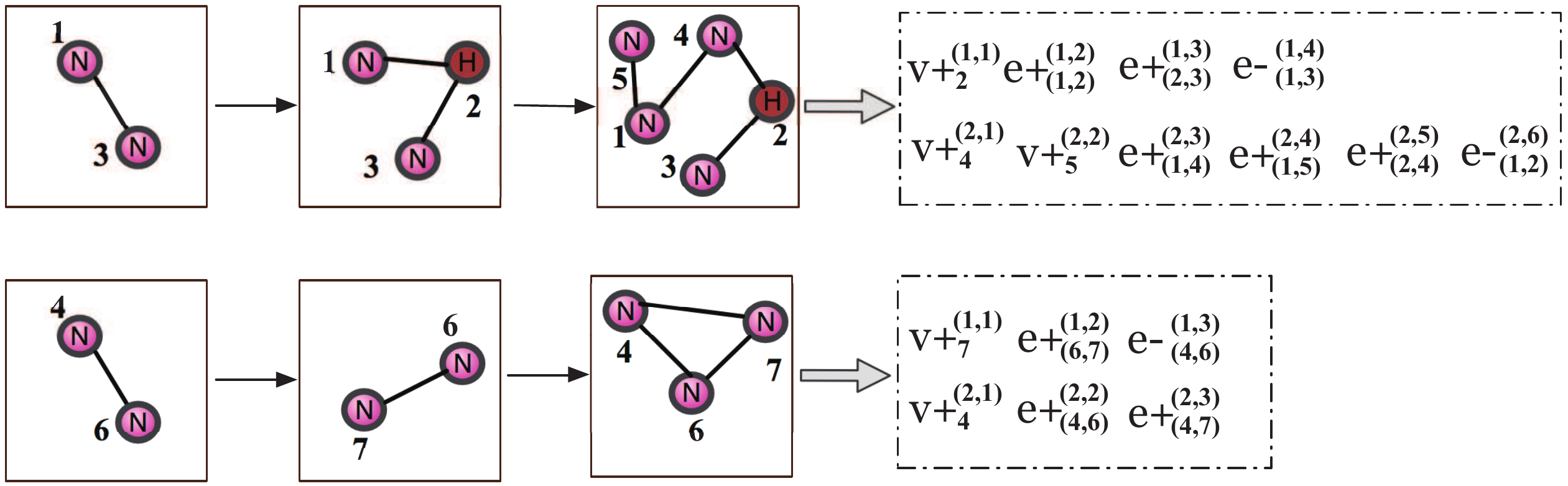}
  \caption{Two graph transformation sequences extracted from the MIT phone call time-variant graph data. The symbol ``N'' represents a normal (non-hub) person and ``H'' represents a hub person. The first graph sequence shows weekly phone call time-variant graph data contain a hub person, while the second graph sequence shows all the participants as normal persons.}\label{fig:experimentExample}
\end{figure}

\begin{figure*}[!t]
  \centering
  \includegraphics[width=0.9\textwidth]{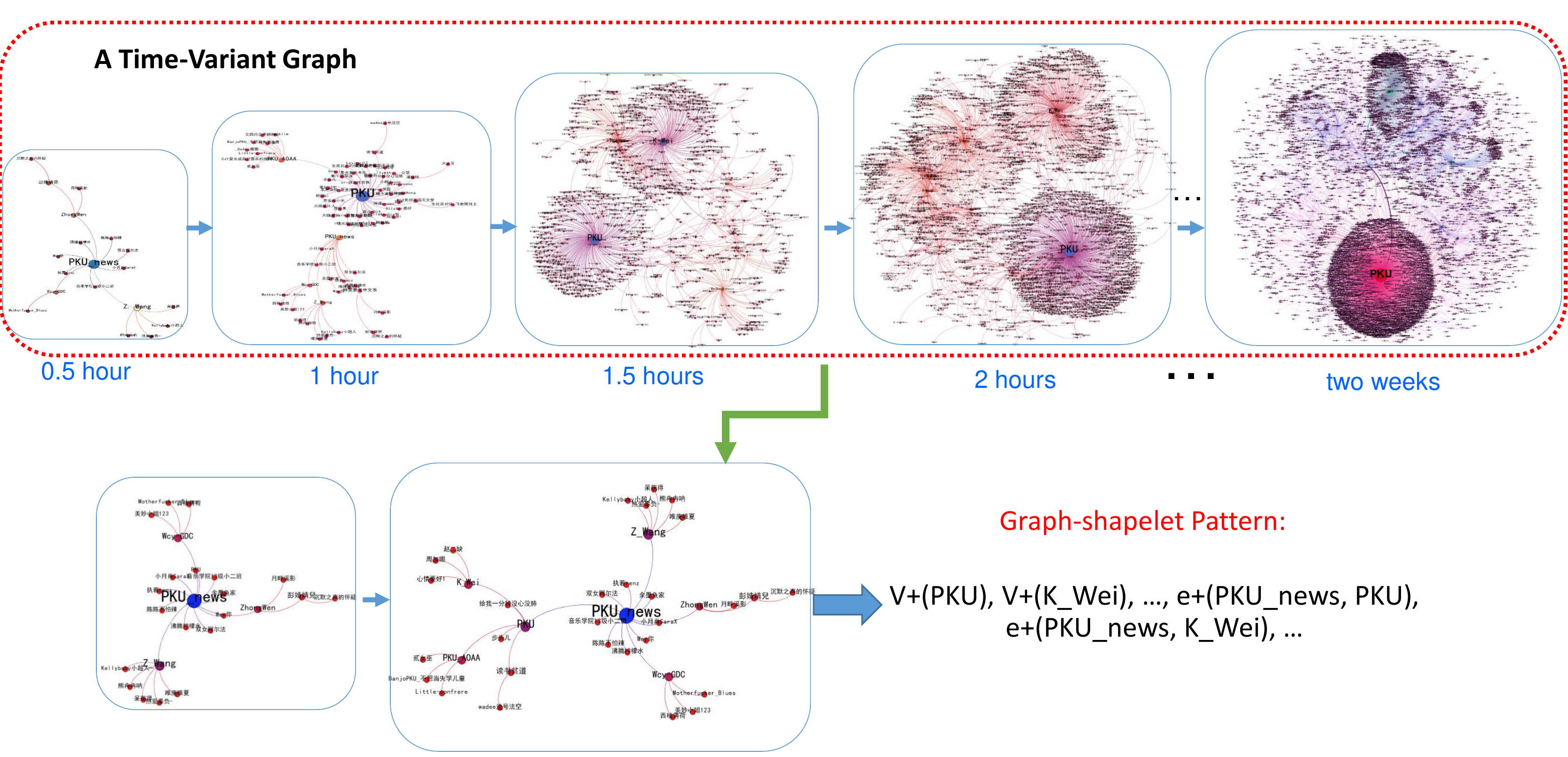}
  \caption{An example of a message propagated in a Sina Weibo time-variant graph. At each propagation stage, the diffusion (including reached vertices and propagation edges) constitutes a graph. The propagation of the message in temporal order will form a set of temporally related graphs. Each Weibo propagation is regraded as a time-variant graph.}\label{fig:weibopropagation}
\end{figure*}

\subsubsection{Efficiency Results}

Fig. \ref{fig:runningtime} shows the running time of the methods, where a subgraph-based method has the lowest running time while a subsequence-based method is the most time consuming. The reason is that subsequence-base method needs to search the entire time-variant graphs to find discriminative substructures whereas the subgraph-based method only mines frequent subgraphs from a single graph. There is no significant difference among \emph{gShapelet}, \emph{gShapelet\_n} and \emph{gShapelet\_e}, because the only difference in these three methods is when transforming a time-variant graph into a univariate time series.

The classification accuracy and running time results show that the graph-shapelet pattern based method has higher accuracy and lower running time than other methods. This is because graph-shapelet pattern based methods consider the transformation relationships between changing graphs, and graph-shapelet patterns are able to learn the discriminative structures of data belonging to different classes. In addition, graph-shapelet patterns are usually much shorter than the original time series, which avoids comparisons being made on the entire data set and results in more efficient running time performance.

\subsubsection{Analysis of gShapelet Algorithm}
Fig.~\ref{fig:lentest} shows the results of gShapelet using both vertices and edges as the graph statistics, respectively compared to gShapelet\_n only using vertices as the graph statistics and gShapelet\_e only using edges as the graph statistics, with respect to various graph-shapelet pattern lengths. We observe that the highest classification accuracy is achieved when \emph{len} is 6, 4, 6 for the Synthetic, MIT, and Weibo time-variant graph data, respectively.

Fig.~\ref{fig:experimentExample} shows a simple yet interesting example extracted from the MIT phone call time-variant graph data. In the first sequence, ``normal'' (non-hub) persons 1 and 2 connect with hub person 2, which leads to an increased number of participants. On the other hand, communications generated by ``normal'' persons have a relatively stable number of users. This implies that the first sequence is likely to contain interesting information, and the potential graph-shapelet patterns can be explored from the corresponding graph transformation sequence.

\section{Discussion} \label{sec:dis}
Intuitively, since the outbreak of an information propagation graph is justified by the number of vertices influenced by the propagation, alert readers may wonder why we transfer time variant-graphs as transformation sequences and further identify graph-shapelet patterns for classification, or why we do not directly use the change in the number of vertices as features for classification. Our hypothesis is that outbreak in information propagation is driven by some special graph structure, rather than simply being determined by vertices which have a large number of degrees. Indeed, an information propagation may not be outbreak even if it reaches some vertices with very high degrees (\textit{i.e.} popular users). This is consistent with the research in~\cite{Leskovec:SDM07}, which observe that information propagation is not mainly spread by vertices with a large number of degrees; rather, it is spread by vertices with a medium number of degrees.

To further validate our hypothesis, we use a real-world case study from Weibo time-variant graph data to demonstrate the discovered graph-shapelet patterns. Fig.~\ref{fig:weibopropagation} is an example of outbreak information propagation which was crawled every half an hour. The Weibo time-variant graph information propagation concerns a news message propagated in a University community regarding an interesting institutional documentary. At different stages, the propagation of the news forms variant graphs. The diffusion of one message in a propagation stage constitutes one time-variant graph, and the diffusion of messages in the community forms a time-variant graph dataset. The message diffusion includes outbreak and non-outbreak time-variant graphs, and our aim is to classify each information propagation graph record into the correct class (outbreak vs. non-outbreak). A naive method of predicting time-variant graph outbreak is to count the volume of vertices, but the time-variant graph is not only influenced by the vertices but by the structures connecting the vertices, such as who is sending the message, who resends the message, when the message is sent or resent, etc., which is precisely captured by time-variant graphs. The general graph classification uses each graph separately in a certain period to discover features. In reality, a diffusion graph will influence subsequent graphs. For time-variant graph classification, we consider the variation between different graphs. Because a graph-shapelet pattern denotes the transformation of vertices and edges between different graphs, it can inherently capture the vertex information, structure information, and temporal variation graphs, for time-variant graph classification.

Fig.~\ref{fig:weibopropagation} shows an example of the mined graph-shapelet patterns. This graph-shapelet pattern example represents the outbreak class and demonstrates that a time-variant graph tends to be outbreak if there is a sequence of special vertex and edge insertions, such as vertex PKU and edge (PKU\_news, PKU) which represents Peiking Universality.

\section{Conclusion}
\label{con}
In this paper, we formulated a new time-variant graph classification task and proposed the use of graph-shapelet patterns as features to represent time-variant graphs for learning and classification. We argued that existing graph classification methods are all based on static settings, where training graphs are assumed to be independent of one another and the structure of each graph remains unchanged. In reality, many applications involve data with dynamic changing structures. Accordingly, a time-variant graph can be used to represent a sequence of graphs and capture the dynamic evolution of the underlying object. The inherent structural dependency and temporal correlations of the time-variant graphs raise significant challenges, and we advocated a graph-shapelet pattern concept to detect significant changes in the time-variant graph as features. By using graph-shapelet patterns to convert a time-variant graph as a tokenized sequence, we can effectively calculate the distance between two time-variant graphs for classification. Experiments on synthetic and real-world data demonstrated that our method is much more accurate than traditional subgraph feature based approaches.

The key innovation of the paper compared to existing work in the area is threefold: (1) a new time-variant graph learning task to model dynamic changes in structured data; (2) a unique graph-shapelet pattern as a feature for capturing the structural and temporal dependency of graphs; and (3) a fast time-variant graph classification model using structure features and edit similarity.

Although our proposed graph-shapelet features and algorithms for classifying time-variant graphs achieve faster running time and high accuracy, some limitations exist that need improvement in the future work: 1) the approach of converting a time-variant graph to a time-series has rooms for improvement; and 2) the proposed algorithm, based on edit similarity, is a straightforward solution for graph classification problems, however more state-of-the-art graph classification algorithms could be applied to achieve a better and faster performance, such as graph boosting (gboost) method.
This work inspires some interesting directions for future work: 1) the problem could be further extended by considering the graph structure when converting a time-variant graph to a time series; and 2) correlations of subgraphs could assist graph stream classification or incremental subgraph features.

\section*{ACKNOWLEDGEMENT}
This work was supported by Australian Research Council Discovery Projects (Nos. DP140102206 and DP140100545).

\bibliographystyle{IEEEtran}
\bibliography{mybibfile}

\end{document}